\documentclass[a4paper,11pt]{article}
\usepackage{jcappub} % for details on the use of the package, please see the JINST-author-manual
\title{\boldmath The MAGPI Survey: co-evolution of baryons and dark matter in star-forming disk-like galaxies at $ 0.1 \lesssim z \lesssim 0.85$ }

\author[1,2,3,4,5,*]{Gauri Sharma, \note{Corresponding author: gauri.sharma@astro.unistra.fr}}
\author[6,7,8]{Andrew J. Battisti,}
\author[7,8]{Emily Wisnioski,} %\author[a,2]{T. Hird\note{Also at Some University.}}
\author[7,8]{J. Trevor Mendel,}
\author[6]{Sabine Bellstedt,}
\author[6,8]{Claudia Del P. Lagos,}
\author[10,8]{Caroline Foster,}
\author[9]{Adriano Poci,}
\author[12, 18, 19]{Katherine E. Harborne,}
\author[10,8]{Ryan Bagge,}
\author[11,8]{Stefania Barsanti,}
\author[11]{Joss Bland-Hawthorn }
\author[13]{Iris Breda,}
\author[11]{Scott M. Croom}
\author[20,8]{Karl Glazebrook,}
\author[15, 16, 8]{Yifan Mai,}
\author[14,8]{Sarah M. Sweet,}
\author[13]{Sabine Thater,}
\author[17]{Lucas M. Valenzuela,}
\author[13]{Glenn van de Ven,}
\author[21]{Sukyoung Yi,}
\author[14]{Tayyaba Zafar,}
\author[13]{and Bodo Ziegler,}
\affiliation[1]{University of Strasbourg, CNRS UMR 7550, Observatoire astronomique de Strasbourg, F-67000 Strasbourg, France}
\affiliation[2]{SISSA– International School for Advanced Studies, Via Bonomea 265, I-34136 Trieste, Italy}
\affiliation[3]{UWC– University of the Western Cape, Department of Physics and Astronomy, Cape Town 7535, South Africa}
\affiliation[4]{IFPU– Institute for Fundamental Physics of the Universe, Via Beirut, 2, 34151 Trieste, Italy}
\affiliation[5]{INFN-Sezione di Trieste, via Valerio 2, I-34127 Trieste, Italy}
\affiliation[6]{International Centre for Radio Astronomy Research, The University of Western Australia, 35 Stirling Highway, Crawley WA 6009, Australia}
\affiliation[7]{Research School of Astronomy and Astrophysics, Australian National University, Cotter Road, Weston Creek, ACT, 2611, Australia}
\affiliation[8]{ARC Centre of Excellence for All Sky Astrophysics in 3 Dimensions (ASTRO 3D) }
\affiliation[9]{Sub-Department of Astrophysics, University of Oxford, Denys Wilkinson Building, Keble Road, Oxford OX1 3RH }
\affiliation[10]{School of Physics, University of New South Wales, Sydney, NSW 2052, Australia}
\affiliation[11]{Sydney Institute for Astronomy, School of Physics, A28, The University of Sydney, NSW 2006, Australia}
\affiliation[12]{Institute for Computational Cosmology, Physics Department, Durham University, South Road, Durham DH1 3LE, UK}
\affiliation[13]{Department of Astrophysics, University of Vienna, T\"urkenschanzstra{\ss}e 17, 1180 Vienna, Austria}
\affiliation[14]{School of Mathematics and Physics, University of Queensland, Brisbane, Qld 4072, Australia}
\affiliation[15]{Australian Astronomical Optics, Macquarie University, Sydney, NSW 2109, Australia}
\affiliation[16]{Astrophysics and Space Technologies Research Centre, Macquarie University, Sydney, NSW 2109, Australia}
\affiliation[17]{Universitäts-Sternwarte, Fakultät für Physik, Ludwig-Maximilians-Universität München, Scheinerstr. 1, 81679 München, Germany }
\affiliation[18]{Centre for Extragalactic Astronomy, Durham University, South Road, Durham DH1 3LE, UK }
\affiliation[19]{Department of Physics, Durham University, South Road, Durham DH1 3LE, UK }
\affiliation[20]{Centre for Astrophysics and Supercomputing, Swinburne University of Technology, P.O. Box 218, Hawthorn, VIC 3122, Australia}
\affiliation[21]{Department of Astronomy and Yonsei University Observatory, Yonsei University, Seoul 03722, Korea}
\usepackage{natbib} %original
\defcitealias{GS23}{GS25}
\defcitealias{GS25}{GS25}
\usepackage{pifont}% http://ctan.org/pkg/pifont
\def \RC{rotation curve}
\def \RCs{rotation curves}
\def \sfgs {star-forming galaxies}

\def \hz{high-$z \ $}

\newcommand{\textblue}[1]{\textblue{#1}}

\abstract{

We present a comprehensive analysis of the dark matter (DM) content and its structural dependence in star-forming disk-like galaxies at intermediate redshifts ($0.1 \lesssim z \lesssim 0.85$), utilizing spatially resolved kinematic data from the Middle Ages Galaxy Properties with Integral Field Spectroscopy (MAGPI) survey. Our sample consists of 266 rotation-supported, main-sequence galaxies. We employ 3DBarolo to model the gas kinematics, obtaining moment maps, surface brightness profiles, rotation velocity profiles, and velocity dispersion profiles. The observed rotation velocity profiles are corrected for inclination and gas pressure gradient to derive intrinsic rotation curves. By modeling these intrinsic rotation curves and decomposing the total dynamical mass profile into the baryonic and the DM components without employing parametric halo models, we estimate the DM fraction ($f_{_{\rm DM}}$) within both the effective radius ($R_{\rm e}$) and the optical radius ($R_{\rm opt}$). We report the following: (1) Low stellar mass galaxies ($M_{\rm star} < 10^{9.5}\, M_\odot$) are strongly DM dominated across all radii, with average $\langle f_{_{\rm DM}} \rangle \sim 0.85$, while high-mass ($M_{\rm star} > 10^{10.5}\, M_\odot$) systems exhibit relatively low DM fractions in their inner regions ($\langle f_{_{\rm DM}} \rangle \sim 0.45$) which is equivalent to local massive disk galaxies (e.g., Milky Way and Andromeda). This suggests a mass-dependent structural dichotomy, most-likely governed by a combination of internal galactic processes and environmental influences. (2) A tight inverse correlation between $f_{_{\rm DM}}$ and baryon mass surface density ($\Sigma_{\rm bar}$), with intrinsic scatter of $\sim 0.12$ dex. This is consistent with an inside-out baryon assembly scenario and suggests that the fundamental structural correlations of galaxies were already established by $z\sim 0.85$. (3) No significant evolution in $f_{_{\rm DM}}$ with redshift across the MAGPI window, and when combined with higher-redshift ($0.6 \leq z \leq 1.5$) data from Sharma et al. 2025, we quantitatively show that the reported decline in $f_{_{\rm DM}}(z)$ is most-likely due to observational biases against low-mass systems at $z > 1$. These results offer empirical evidence for a scenario in which disk-like galaxies evolve through a co-regulated build-up of baryonic and DM components, preserving internal structural regularities (such as the total mass distribution, and rotation-curve shape) throughout cosmic time.
}

\begin{document}
\maketitle
\flushbottom

\section{Introduction}
\label{sec:intro}
Under the current standard model of cosmological structure formation, dark matter (DM)$-$an elusive, non-luminous form of matter$-$is thought to constitute the majority of matter in the Universe \citep[e.g.,][]{Peebles1993}. Despite its prevalence, DM  remains undetected by direct means; its existence is inferred primarily from the gravitational effects it exerts on the visible matter. In particular, the discovery of flat rotation curves in the outer regions of galaxies provided compelling evidence that the visible mass alone is insufficient to explain the observed galactic dynamics \citep{rubin1980, bosma1981}. This discrepancy led to the hypothesis that galaxies are embedded in extended DM  halos that far outweigh their luminous components in mass \citep[e.g.,][]{rubin1980, bosma1981, Sofue2001, PS2019}. Furthermore, cosmological models and simulations within the standard cold DM  ($\Lambda$CDM) framework support the picture that DM  halos act as a gravitational scaffolding for galaxy formation and evolution \citep[][and references therein]{Sims_review_2022}. However, this paradigm faces several persistent small-scale challenges \citep[e.g.,][]{Bullock2017, Sales2022}, including the ``too big to fail'' \citep{Boylan-Kolchin11} and the ``diversity of rotation curves'' \citep{Oman15} problems. These persistent problems highlight that our understanding of DM  and baryon evolution in galaxies is incomplete.

In this context, detailed observations of mass distributions within the  galaxies offer crucial insights. The fraction of mass contributed by DM  versus baryons varies widely with galaxy type, mass, and size, underscoring the crucial role of DM  in galaxy dynamics. In nearby star-forming disk galaxies, DM  typically accounts for roughly 40–50\% of the dynamical mass within the inner region (inside half-light radius $R_{\rm e}$), rising to about 70–90\% within $\sim 3R_{\rm e}$ which encompasses most of the galaxy’s light \citep{Kassin2006, Martinsson2013, Courteau2015}. In contrast, nearby early-type galaxies (ellipticals and lenticulars) often show much lower DM  contributions in their central regions, sometimes as little as $\lesssim 20\%$ within $R_{\rm e}$ \citep{ATLAS3D_2013}. However, in massive early-type galaxies (with $M_{\rm star} \gtrsim 10^{10}\, M_{\odot}$) the central DM  fraction tends to increase with stellar mass, yet their outskirts still reach similarly high DM dominance  by the radii that encompass most of the stellar light \citep[70–90\% within $\sim5R_{\rm e}$;][]{William2020}. These differences suggest that the assembly history and baryonic processes (e.g., star-formation) of galaxies are intimately connected with the distribution of DM  relative to stars; hence, play a crucial role in shaping galaxy structure.

The rationale is that the baryons are expected to be more centrally concentrated than DM due to their dissipative nature: baryons can cool and condense, whereas collisionless CDM cannot. This naturally leads to galaxy centres becoming more baryon-dominated compared to their outskirts. However, this effect alone does not explain the observed differences in DM fractions between low-mass and massive galaxies. A plausible explanation lies in the interplay between ``downsizing'' and the evolution of gas accretion in a $\Lambda$CDM universe. In the downsizing scenario, supported by local observations \citep[e.g.,][]{Thomas05, Thomas10}, massive galaxies form earlier and more rapidly than their low-mass counterparts. According to tidal torque theory \citep{Catelan96a} and hydrodynamical simulations \citep{El-Badry18, Garrison-Kimmel18}, gas accreted at early times in a $\Lambda$CDM universe has lower specific angular momentum and is more filamentary than gas accreted later \citep[see][for a review]{Lagos20}. Such filamentary flows can penetrate deep into galactic potential wells, forming highly concentrated progenitors of present-day massive galaxies \citep{Dekel14, Zolotov15, Danovich15, Waterval25}. 

In contrast, low-mass galaxies typically assemble later and undergo more extended star formation histories, making them systematically more susceptible to late-time accretion. Later accreted gas, having gained higher specific angular momentum through tidal torques \citep{Catelan96a, Dutton2011, Brook2011, Stewart2013}, settles at larger radii. However, the retention of extended gas distributions is not exclusive to the low-mass regime; both massive and low-mass disk galaxies exhibit extended neutral gas (HI) reservoirs. At a fixed stellar mass, a galaxy's gas fraction correlates strongly with its specific angular momentum, meaning that more gas-rich systems naturally develop higher specific angular momentum and larger physical sizes \citep[e.g.,][]{Lagos2017, Zoldan2018, ManceraPina2021}. Ultimately, whether driven by the delayed, late-time assembly of low-mass systems or the general retention of gas-rich, extended disks, this extended settling encompasses a greater fraction of the halo's DM relative to the galaxy's stellar size. Regardless of the details, the specific mechanisms of baryonic accretion and assembly are fundamental in dictating how centrally concentrated the baryons end up being relative to the DM.

In addition to the effects of assembly and accretion history, a further layer of complexity arises from baryonic feedback processes. Intense star formation, supernovae, and active galactic nucleus (AGN) activity can drive powerful gas outflows, which alter the gravitational potential and redistribute both the baryons and DM. Some simulations show that repeated feedback episodes can transform initially steep DM  cusps into lower-density cores, reducing the DM  fraction in the centers of galaxies over several gigayears \citep{Pontzen2012, Pontzen2014, Dutton2016, El-Zant2016, JF2020a, Dekel2021, Li2023}. Thus, the final distribution of DM  and baryons is shaped by a combination of how baryons are accreted, assembled, and subsequently redistributed. Measuring the radial distribution of DM  and its evolution with cosmic time is therefore crucial; it not only constrains the fundamental nature of DM  but also provides critical tests of galaxy formation theories and feedback models in cosmological simulations.

% Paragraph : the new advances in IFUs and current studies 
In the past decade, advances in integral field unit (IFU) surveys have opened new avenues for studying galaxies’ internal kinematics out to high redshift. 
IFU observations now enable the construction of spatially resolved observations (e.g., velocity maps and rotation curves) of galaxies up to $z\sim 6$ \citep[e.g.,][]{Forster2020, Fujimoto2024, Ubler2024}, allowing direct probes of the distribution of baryonic and dark mass in distant galaxies. Early IFU studies of high-$z$ star-forming galaxies (SFGs) yielded surprising results. For example, \cite{Genzel2017} analysed a sample of 7 high-redshift ($z \approx 0.6 - 2.6$) SFGs and \cite{Lang2017} extended this to 101 systems using staking method, finding that high-redshift (hereafter high-$z$) galaxies show declining outer rotation curves. Such a decline is uncommon among typical local star-forming disk galaxies (which generally have flat rotation curves), being observed locally only in a few extremely massive, high surface-brightness disks \citep[e.g.,][]{rubin1980, PS1996}. Genzel et al. and Lang et al. argued that the falling rotation curves (beyond $1.5 R_e$) at high-$z$ could be explained by a combination of very high baryon fractions in the galaxy centers and significant pressure support from turbulent gas. Similarly, other high-$z$ studies of both late-type and early-type galaxies have reported relatively low DM  fractions within the $R_{\rm e}$ \citep[e.g.,][]{Burkert2016, Wuyts2016, Price2016, Ubler2018, Genzel2020, Mendel2020, Price2021, Genzel2022}.

% Paragraph: Sharma et al. previous papers 
However, results on the DM content in high-$z$ \sfgs\ have not yet converged. For example, \cite{AT2019} examined the \textit{stacked} (averaged) rotation curves of $\sim$1500 SFGs across $0.6 \lesssim z \lesssim 2.2$. They found that stacked rotation curves are flat out to large radii, closely resembling those of local SFGs. Moreover, their analysis indicated substantial DM contributions: more than 50\% of the total mass within $3.5\, R_{\rm e}$, in-line with typical local disk galaxies \citep{PS1996, Martinsson2013, Courteau2015}. In another study \cite{GS21a} applied 3D forward modelling to data from the KMOS Redshift One Spectroscopic Survey (KROSS; \cite{sharples_2014}, \cite{stott2016}) to derive \textit{individual} rotation and velocity dispersion curves for galaxies at $z \sim 1$. By correcting for beam-smearing and pressure support, they recovered the intrinsic rotation curves of over 200 rotation-supported disks, finding that most of the \RCs\ are flat out to $3 R_{\rm e}$. In \cite{GS21b}, the authors used the \cite{GS21a} intrinsic rotation curves and employed a halo model independent method to estimate the DM content. They showed that galaxies are DM dominated across the galactic scales. 

Extending this analysis, \cite[][hereafter GS25]{GS23} applied the same methodology as \cite{GS21a, GS21b} to a sample of 263 individual rotation-supported SFGs spanning $0.6 \leq z < 2.5$, including the above KMOS samples, the largest high-$z$ sample with resolved kinematics to-date. They found that the majority ($\sim 75\%$) of these galaxies have DM dominated outer disks, with DM  contributing between $50 - 90\%$ of the total mass within $R_{\rm e}$ out to $R_{\rm out}$.\footnote{In general, the scale length (or characteristic radius) are associated with various quantities that decrease exponentially such as the surface brightness. The stellar disk edge is defined as $3.2 \ R_{\rm D}$ ( $=$ $1.89 \ R_{\rm e}$), where the stellar surface luminosity $\propto \exp(-r/R_{\rm D})$. Therefore, disk scale lengths in terms of the effective radius: $R_{\rm D} = 0.59 \ R_{\rm e}$. Based on studies of local disk galaxies \citep{PS1996}, the radius enclosing 80\% of the stellar light is called optical radius $R_{\rm opt} = 1.89 \, R_{\rm e}$, while the outer radius ($R_{\rm out}$), corresponding to 99\% of the stellar light, is $R_{\rm out} = 2.95 \, R_{\rm e}$ \citep{GS21a}.} A recent study by \cite{Ciocan2025} independently confirm these findings using a halo-model dependent approach, further solidifying the evidence that high-$z$ disks generally have massive DM  halos, see also \cite{Pelliccia2017, H17, Bouche2022, Ubler2024, ManceraPina2026}. Although there is growing evidence that support the flat rotation curves and DM dominated galaxies at high-$z$, limitations such as a small sample size, heterogeneous dynamical modelling techniques, and lack of resolved kinematics at intermediate redshifts ($0.1 \leq z < 0.6$) have prevented us from drawing a coherent picture of DM and baryon co-evolution over cosmic time. 

In this work, we address the aforementioned limitations. We present a comprehensive study of DM fraction at the intermediate-redshift range ($0.1 \lesssim z \lesssim 0.85$),  utilizing spatially resolved kinematics from the Middle Ages Galaxy Properties with Integral Field Spectroscopy (MAGPI) survey \citep{Foster2021}. It is a Large Program (Program ID: 1104.B-0536) using Multi-Unit Spectroscopic Explorer (MUSE) on the Very Large Telescope (VLT) at European Southern Observatory (ESO), delivering high-resolution spectroscopy of both stars and ionized gas at a median redshift of $\langle z \rangle \sim 0.35$ (corresponding to look-back times of $3-4$ Gyr). We provide the details of the survey and dataset in Section~\ref{sec:data}. In particular, we will employ the same methodology as used by \citetalias{GS23}, which is briefly outlined in Section~\ref{sec:Barolo} \& \ref{sec:method}. We present and discuss the results in Section~\ref{sec:result}. Finally, we combine the MAGPI dataset with the \citetalias{GS23} sample, a statistically robust sample of 529 star-forming galaxies, enabling quantitative analysis of baryon and DM co-evolution as a function of galaxy mass, spatial scale, and redshift, which is discussed in Section~\ref{sec:FdmEvo}. A summary of the main findings is provided in Section~\ref{sec:summary}. Throughout this analysis, we assume a flat $\Lambda$CDM  cosmology with $\Omega_{m,0} = 0.27$, $\Omega_{\Lambda,0} = 0.73$, and $H_0 = 70 \ \mathrm{km \ s^{-1} \ Mpc^{-1}}$.

%%%%%%%%%%%%%%%%%%%%%%%%%%%%%%%%%%%%%%%%%%
\section{Dataset}
\label{sec:data}
Using the MUSE on the VLT, the MAGPI survey is designed to investigate the spatially resolved stellar and ionized gas properties of galaxies at lookback times of $3$–$4$ Gyr ($\langle z \rangle \approx 0.35$). The survey targets 60 primary galaxies at $z \sim 0.3$ with stellar masses $M_{\rm star} > 7 \times 10^{10} \, M_\odot$, along with $\sim 100$ satellite galaxies spanning a range of environments—from isolated systems to galaxies in groups and clusters. Of these, 56 primary targets were selected from the GAMA survey \citep{Driver2011, Driver2022}, with the remaining four drawn from archival MUSE observations of the Abell 370 \citep{Abell370, Lagattuta2017} and Abell 2744 \citep{Abell2744, Mahler2018} fields.

Observations were conducted using the MUSE Wide Field Mode, covering a wavelength range of $4650$–$9300\, \text{\AA}$ with a spectral sampling of $1.25 \, \text{\AA}$. Each MAGPI pointing spans a $1^{\prime} \times 1^{\prime}$ field of view centered on a primary target. The spatial sampling is $0.2^{\prime\prime}$ per pixel, and the typical point spread function (PSF) full width at half maximum (FWHM) achieved in the $R$-band is $\sim 0.55^{\prime\prime}$. Each field was observed in six observing blocks, each comprising two exposures of $1320 \, \mathrm{sec}$, totaling $4.4$ hours per field and $246$ hours of on-source time for the full survey. Observations employed a ground-layer adaptive optics (GLAO) system to enhance image quality, resulting in a spectral gap between $5780$–$6050 \, \text{\AA}$ due to the sodium laser notch filter of the GALACSI system \citep{Hartke2020}.

A detailed description of the data reduction will be provided in Mendel et al. (in preparation). Briefly, reduction is based on the MUSE-pipeline \href{https://github.com/emsellem/pymusepipe}{\texttt{PY-MUSEPIPE}} \citep{Weilbacher2020}, with sky subtraction performed using the Zurich Atmosphere Purge (ZAP) software \citep{Soto2016}. For each detected source, a ‘minicube’ is extracted from the full MUSE datacube, centered on the galaxy and encompassing the maximum extent of its `dilated' segmentation map, as determined by \texttt{PROFOUND} \citep{Robotham2018}. All MUSE extensions—PRIMARY, DATA, STAT, LSF, PSF, EXPOSURE—are retained, and two mask extensions are added to encode the dilated and undilated segments (MASK, MASKUNDILATED).\footnote{The software \href{https://github.com/musevlt/mpdaf}{\texttt{MPDAF}} is used to produce the minicubes.} Spectral modelling of the `minicubes' is performed by Battisti et al. (in preparation) using \href{https://pypi.org/project/gistPipeline/}{\texttt{GIST v3.0}} \citep{Bittner2019}, with outputs resampled back onto the native MUSE wavelength grid. The resulting data cubes preserve the $[x,y,z]$ structure and include best-fit stellar continuum, emission line models, and residual spectra (minicube – [continuum + emission]). 

Morphological parameters such as axis-ratios ($b/a$) and effective radius ($R_{\rm e}$) are derived from 2D surface brightness modelling using \href{https://github.com/ICRAR/ProFit}{\texttt{ProFit}} \citep{Robotham2017}, assuming a single S$\acute{e}$rsic component. The fitting is done in the MAGPI i-band (i.e. on a mock i-band image computed from the respective MAGPI datacube) using the \texttt{scipy.minimize} function and the `Nelder-Mead' method, which explicitly convolves the model with the instrumental PSF during the fitting process. We utilize a single-component fit here primarily to obtain robust measurements of the global effective radius ($R_e$) and inclination. A full multi-component photometric decomposition (disk + bulge) is highly degenerate at the spatial resolution of our ground-based data as shown in various studies such as \cite{Haussler2007,Peng2010,Simard2011}. Instead, we use this robust global $R_e$ to anchor the empirical scaling relations that define the separate disk and bulge potentials in our dynamical mass modeling (see Section~\ref{sec:MModel}).
To obtain the inclination angle ($\theta_i$), we use these axis-ratios and convert it into inclination according to: 
	\begin{equation}
		\centering
		cos^2 \theta_{i} = \frac{(b/a)^2 - q_0^2}{1-q^2_0}
	\end{equation}
where $q_0$ is the intrinsic axial ratio of an edge-on galaxy (e.g. \cite{TFR1977}), which could in principle have values in the range $0.1$ – $0.65$ (e.g. \cite{Weijmans2014}); here, we use the commonly assumed value $q_0 = 0.2$, which is applicable for thick discs commonly found at \hz \citep{H17}; this value is also often adopted at low-$z$ \citep[e.g.,][]{Padilla2008, Unterborn2008, Rodriguez2013}. Spectroscopic redshifts are estimated using Manual and Automatic Redshifting software, \href{https://matteofox.github.io/Marz/}{\texttt{MARZ}} \citep{Hinton2016}.
%
%Padilla & Strauss 2008; Unterborn & Ryden 2008; Rodríguez & Padilla 2013
%
%
\\
\\  
\textit{Star-formation rates:}
We adopt integrated star-formation rates (SFRs) estimated from 1D Balmer line flux ratio measurements extracted within the dilated segmentation masks (Battisti et al., in preparation). Observed deviations from these intrinsic line ratios can be attributed to dust reddening \citep{Osterbrock2006}, with shorter wavelengths being more absorbed than longer wavelengths (e.g., $H_\beta$ appears fainter relative to $H_\alpha$ than the expected ratio). For the intrinsic line ratios, we assume Case-B recombination, with $T_e =  10^4 K$ and $n_e=100 \ cm^3$ \citep{Osterbrock2006}. Dust attenuation is corrected using the Milky Way extinction curve from \cite{Fitzpatrick2019}. The effect of dust reddening is stronger for lines that are more separated in wavelength, and the uncertainties on dust corrections are generally larger when using line ratios that are closer in wavelength (e.g., $H_\gamma/H_\delta$ instead of $H_\alpha/H_\beta$). This is factored into the SFR uncertainties. Finally, line ratios fluxes are converted to SFRs following the calibration of \cite[][$SFR=5.5\times 10^{-42} L_{H\alpha}$]{Calzetti2013}, which assumes a \cite{Kroupa2013} initial mass function (IMF). When Halpha is not available ($z>0.4$), we use dust corrected $H_\beta$ or other emission lines to estimate SFR by adopting the assumed case-B recombination intrinsic line ratios (e.g., $F_{H_\alpha} = 2.86\ F_{H_\beta}$).
%
%\footnote{The Balmer lines $H\alpha$, $H\beta$, $H\gamma$, and $H\delta$ are centered at $6562.8\,\text{\AA}$, $4861.3\, \text{\AA}$, $4340.5\, \text{\AA}$, and $4104.7\,\text{\AA}$, respectively.}
\\
\\
\textit{Stellar masses:} %
Stellar masses are derived from broad-band photometry (of ugriZYJHK$_s$, bands) by fitting the spectral energy distributions (SEDs) using the \texttt{ProSpect} code \citep{Robotham2017}. Briefly, a high signal-to-noise detection image is utilized to generate source segmentation maps via a watershed de-blending algorithm. These defined segments are subsequently applied across all available broad-band filters to extract matched-aperture photometry. \textit{ProFound} dynamically handles segment dilations to account for varying point spread functions across different observing bands, ensuring that a consistent physical footprint is measured across the full spectral energy distribution. The fits utilize stellar population synthesis models from \cite{Bruzual2003}. Dust attenuation of stellar light is modeled following the two-component prescription of \cite{Charlot_Fall_2000}, and dust emission is implemented using the dust templates of \cite{Dale2014}. For full methodological details, we refer the reader to \cite{Bellstedt2020}.

We compared SFRs for the full sample derived from SED fitting with those estimated from Balmer emission lines and found a systematic offset of approximately 0.07~dex. This offset is propagated when estimating the uncertainties on the dark matter fraction. In general, broadband SED-based SFRs are higher than those derived from Balmer decrement measurements; however, for high-stellar-mass galaxies ($M_{star}\geq 10^{10.5}\ M_\odot$), the SED-based SFRs can be up to two orders of magnitude lower than the Balmer-derived values. It should be noted that SED-derived SFRs are intrinsically more uncertain, as the fits are based solely on \textit{ugriZYJHK\(_s\)} broadband photometry and lack far-infrared (FIR) coverage, which limits constraints on dust attenuation.

\begin{figure*}
	\begin{center}
\includegraphics[angle=0,height=8.0truecm,width=13.0truecm]{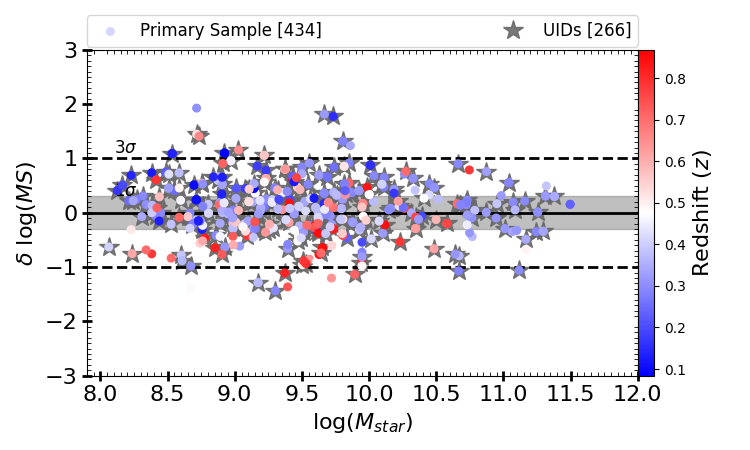} 
\caption{Offset from the star-forming main sequence (MS) as a function of stellar mass ($M_{\rm star}$). The vertical axis shows the logarithmic deviation from the MS, defined as $\delta\log(\mathrm{MS}) = \log(\mathrm{SFR} / \mathrm{SFR}_{\rm MS;}(z, M{\rm star}))$, where $\mathrm{SFR}_{\rm MS}$ is the redshift- and mass-dependent main sequence relation from \cite{speagle14}. Full primary MAGPI sample ($N=434$) sample is represented by circles color coded for redshift, and the gray stars highlight the number of Unique IDs selected after applying kinematic modelling quality cuts. The gray shaded area represents the $1\sigma$ scatter of \cite{speagle14} MS relation. The majority of galaxies lie within the $\pm 2-3\sigma$ range of the \cite{speagle14} MS relation, i.e., final sample is consistent with a population of star-forming galaxies.}
\label{fig:MSequence}
\end{center}
\end{figure*}

\begin{figure*}
	\begin{center}
\includegraphics[angle=0,height=11.0truecm,width=15.5truecm]{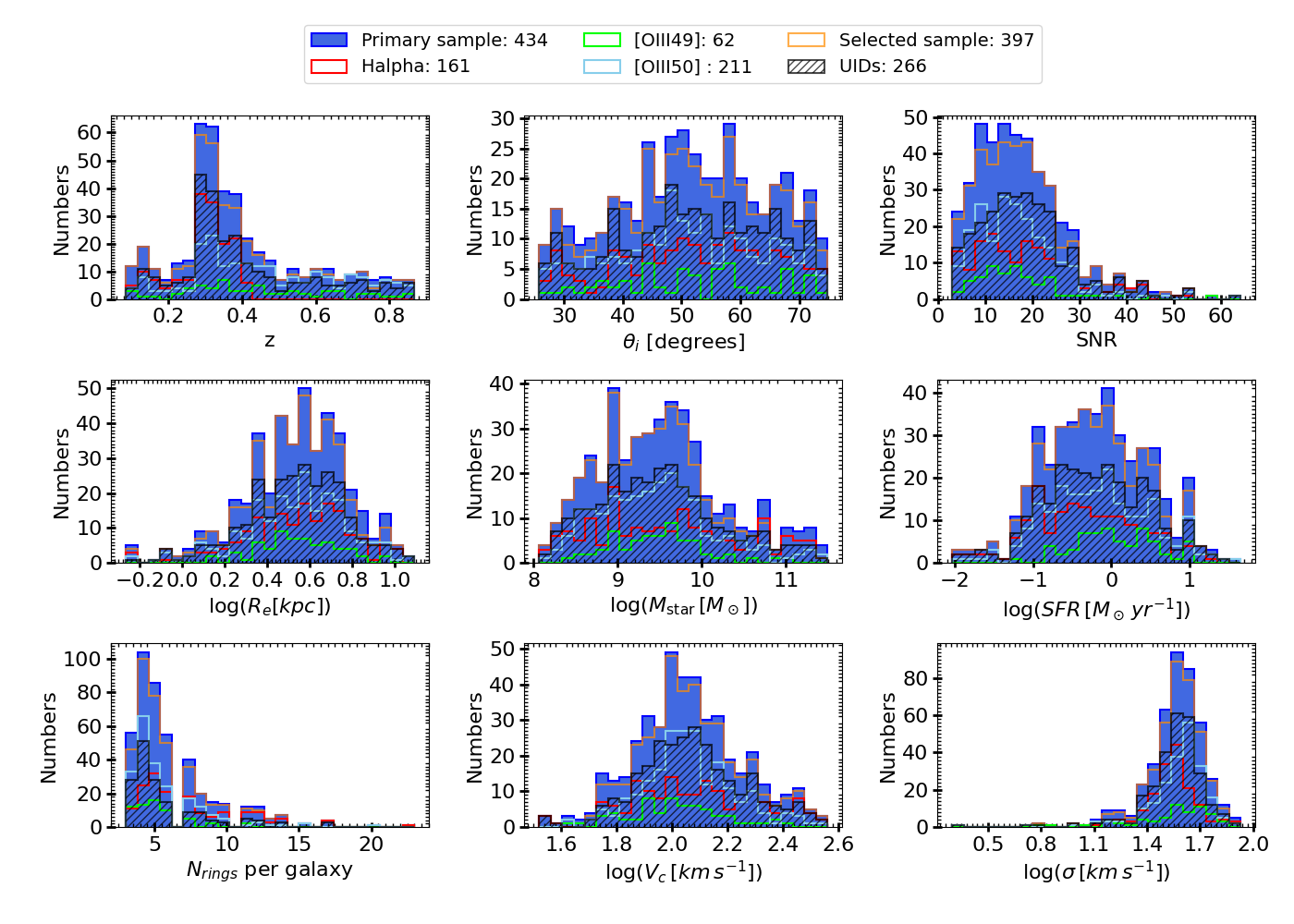} 
\caption{Distributions of key galaxy properties for the full sample used in this study. Each panel shows histograms for the primary sample (blue; 434 galaxies) and subsets based on available emission line measurements: H$\alpha$ (red; 161), [OIII]$\lambda\, 4969\,\text{\AA}$ (green; 62), and [OIII]$\lambda\, 5008\,\text{\AA}$ ( sky-blue; 211). The kinematically selected sample (397) shown by orange histograms, and the black hatched histogram shows the Unique IDs (266) in this sample. Panels show (from left to right, top to bottom): redshift ($z$), inclination angle ($\theta_i$; degrees), integrated signal-to-noise ratio (SNR), effective radius $R_e$ (kpc), stellar mass ($M_{star}$; $M_\odot$), star formation rate (SFR; $M_\odot\,{\rm yr}^{-1}$), number of resolution elements ($N_{rings}$ per galaxy), circular velocity computed at $R_{opt}$ ($V_c;\ km\sec^{-1}$), and velocity dispersion ($\sigma;\ km\sec^{-1}$). The final kinematic sample and UIDs are representative of the primary sample in all fundamental properties, confirming the absence of any selection biases in the working sample.}
\label{fig:MAGPI-hist}
\end{center}
\end{figure*}

\subsection{Sample selection}
\label{sec:sample}
We utilize data from 53 out of 60 observed MAGPI fields, comprising minicubes for approximately 2600 galaxies spanning the redshift range $z \approx 0.1$–$0.85$. Our primary selection includes galaxies with confirmed spectroscopic redshifts. From this set, we identify galaxies with \textit{resolved} gas emission in one or more of the following lines: $H\alpha$ ($6563\, \text{\AA}$), [OIII] ($4969\, \text{\AA}$), and [OIII] ($5008\, \text{\AA}$). As discussed in Section~\ref{sec:emline-RCs} and demonstrated in Figure~\ref{fig:emline-RCs}, these different emission lines, despite variations in line strength and ionization potential, consistently trace the same underlying gravitational potential. Therefore, we decided to use multiple emission lines to expand, both, the statistical power of the sample and enhances the reliability of the kinematic and dynamical inferences drawn from the dataset. Hence, a galaxy remains in the work sample if at-least one of these lines are present in the corresponding minicube, and a galaxy is repeated if it has multiple aforementioned emission lines.\footnote{Final measurement of dark matter fraction of repeated galaxies are weighted as discussed in Section~\ref{sec:emline-RCs}.} Stellar continuum is subtracted using spectral fits obtained from the GIST pipeline, after which we calculate the integrated signal-to-noise ratio (SNR), within the square aperture each-side covering 30 spaxels ($6.2^{\prime\prime}$), of the representative emission line around the central wavelength. Galaxies are retained only if their globally integrated emission line exhibits an average $SNR > 3$ within the Gaussian fit. Additionally, we exclude galaxies located at the edge of the minicubes to avoid boundary artifacts.

To mitigate projection effects in the kinematic analysis—such as radial or tangential motions and extinction-induced distortions that leads to a flattened velocity profile—we select only galaxies with intermediate inclinations in the range $25^\circ \leq \theta_i \leq 75^\circ$. We further restrict the stellar mass of our final sample to $10^8 \le M_{\star} [M_{\odot}] \le 10^{11.5}$. This selected sample is referred to as primary sample, with the majority (99\%) of galaxies lying within the $2-3\sigma$ range of the star-forming main sequence (SFMS) defined by \cite{speagle14}, as shown in Figure~\ref{fig:MSequence}. That is, the \textit{primary sample} is chosen such that it, somewhat, excludes dwarf galaxies, as well as low-surface-brightness or ultra-diffuse systems, allowing us to focus on a representative population of star-forming disk-like galaxies (i.e., analogues to local Sb-type systems). While high-resolution dynamical studies in the nearby Universe routinely probe galaxies well below this stellar mass regime \citep[e.g.,][]{Read2016, ManceraPina2026}, our selected range encompasses the bulk of the main-sequence rotation-supported disk-like population. This ensures a robust, like-for-like evolutionary comparison between our intermediate-redshift MAGPI sample and existing high-$z$ datasets.  We note that the final sample is defined following a secondary selection applied after the kinematic modelling, as described in Section~\ref{sec:Fsample}.

In Figure~\ref{fig:MAGPI-hist}, we present the distribution of key physical and morphological properties of our primary sample. Within this primary sample, we select galaxies with brightest lines for kinematic modelling: 161 using $H\alpha$, 62 using [OIII:4969], and 211 using [OIII:5008]. The redshift range of the sample extends from $z = 0.1 - 0.85$, with $H\alpha$ predominantly probing the lower-redshift regime ($z \approx 0.1 - 0.4$), while the [OIII] lines are prominent across whole redshift range. As illustrated in Figure~\ref{fig:MAGPI-hist} and mentioned above, our selection criteria ensure robust coverage in SNR, inclination, and stellar mass. The resulting sample spans a wide range of effective radii ($R_{\rm e} = 0.5$–$10\,{\rm kpc}$) and star-formation rates ($\mathrm{SFR} = 0.01$–$32\, M_\odot\, \mathrm{yr}^{-1}$), making it well-suited for disk galaxy population study at intermediate redshifts. 
%%%%%%%%%%%%%%%%%%%%%%%%%%%%%%%%%%%%%%%%%%%%%

%%%%%%%%%%%%%%%%%%%%%%%%%%%%%%%%%%%%%%%%%%%%%
\begin{figure*}
	\begin{center}
\includegraphics[angle=0,height=4.5truecm,width=14.truecm]{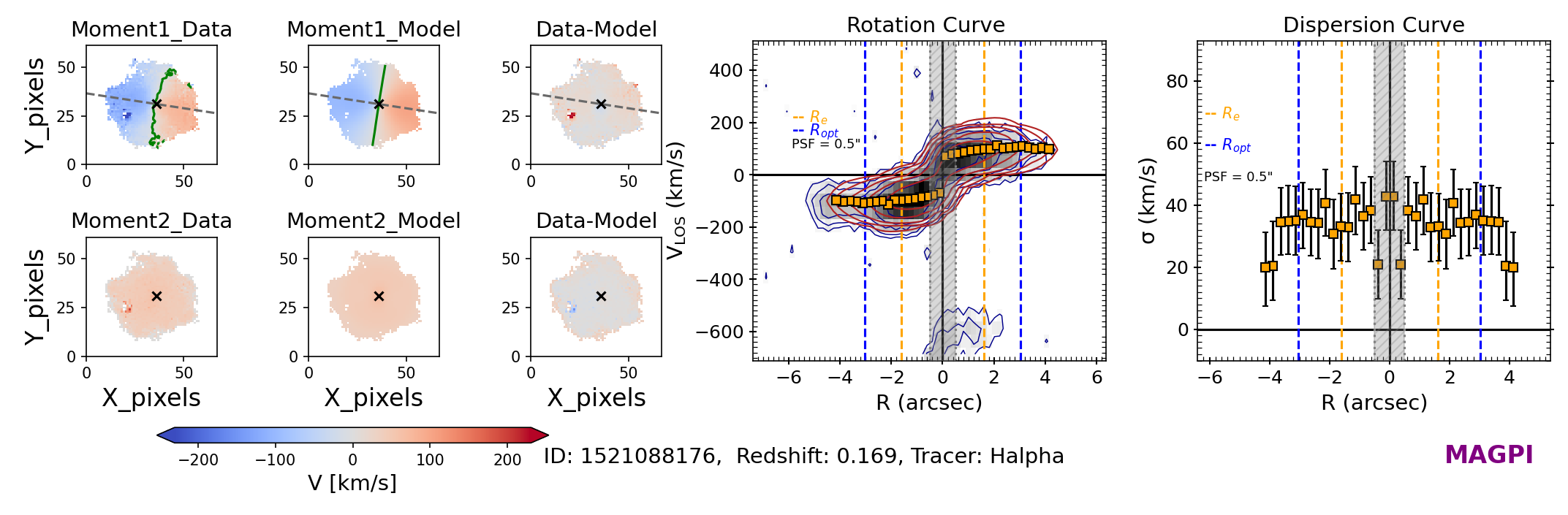}

\includegraphics[angle=0,height=4.5truecm,width=14.0truecm]{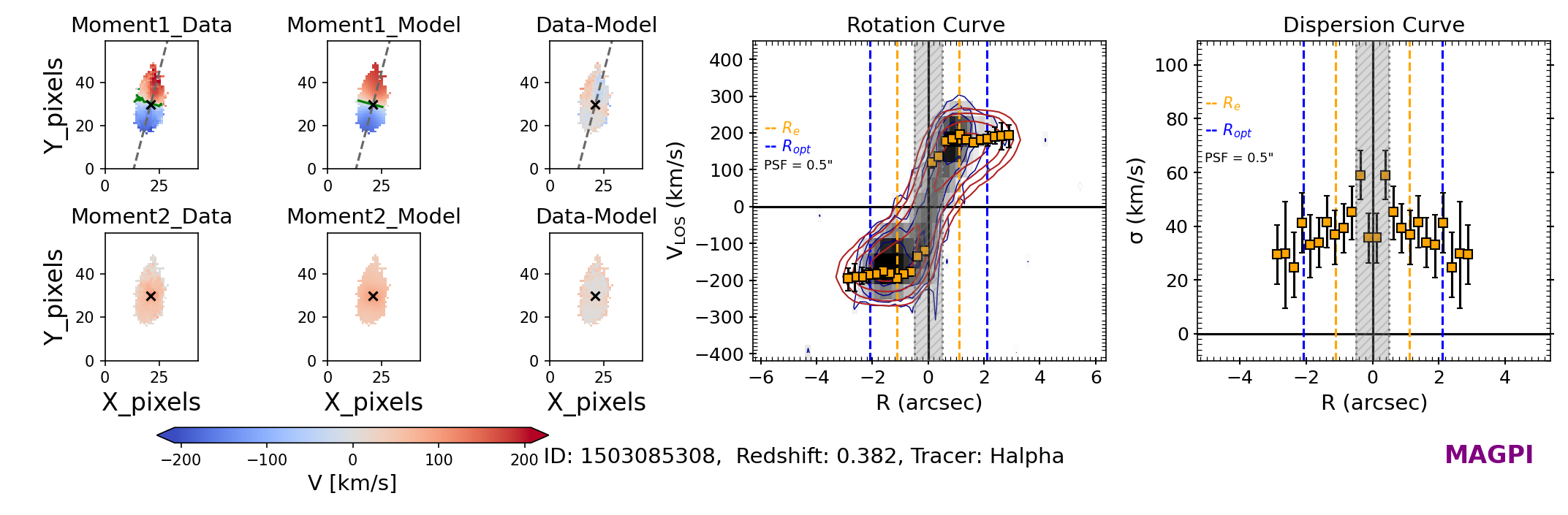} 

\includegraphics[angle=0,height=4.5truecm,width=14.0truecm]{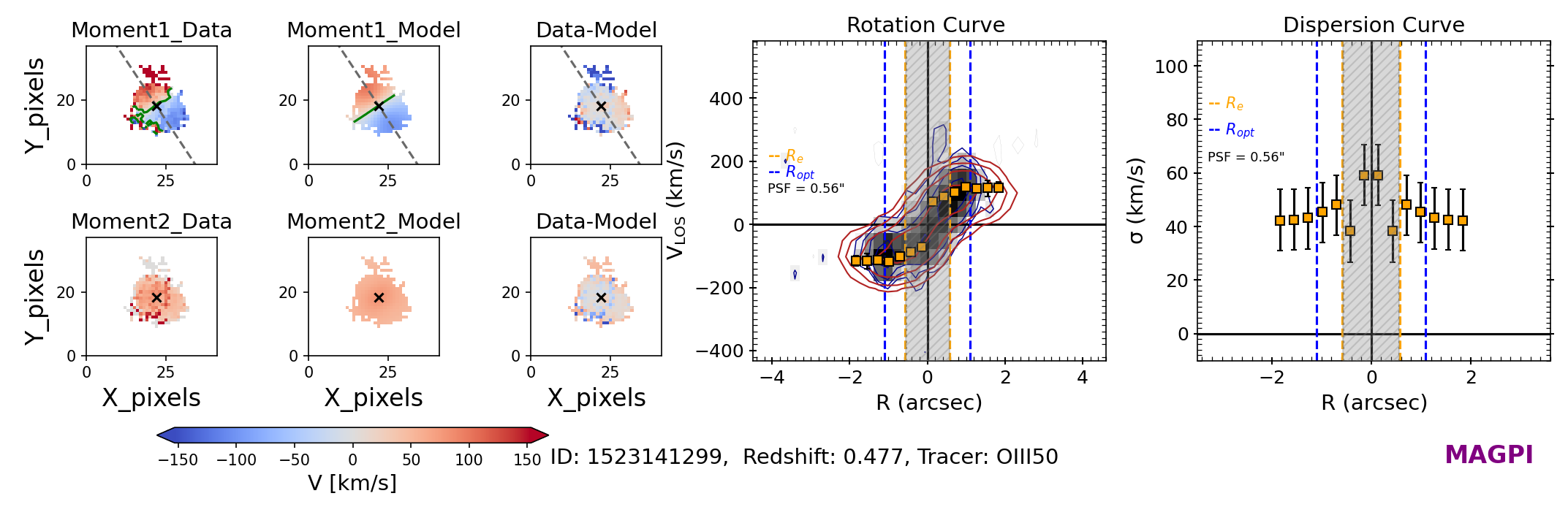}

\includegraphics[angle=0,height=4.5truecm,width=14.truecm]{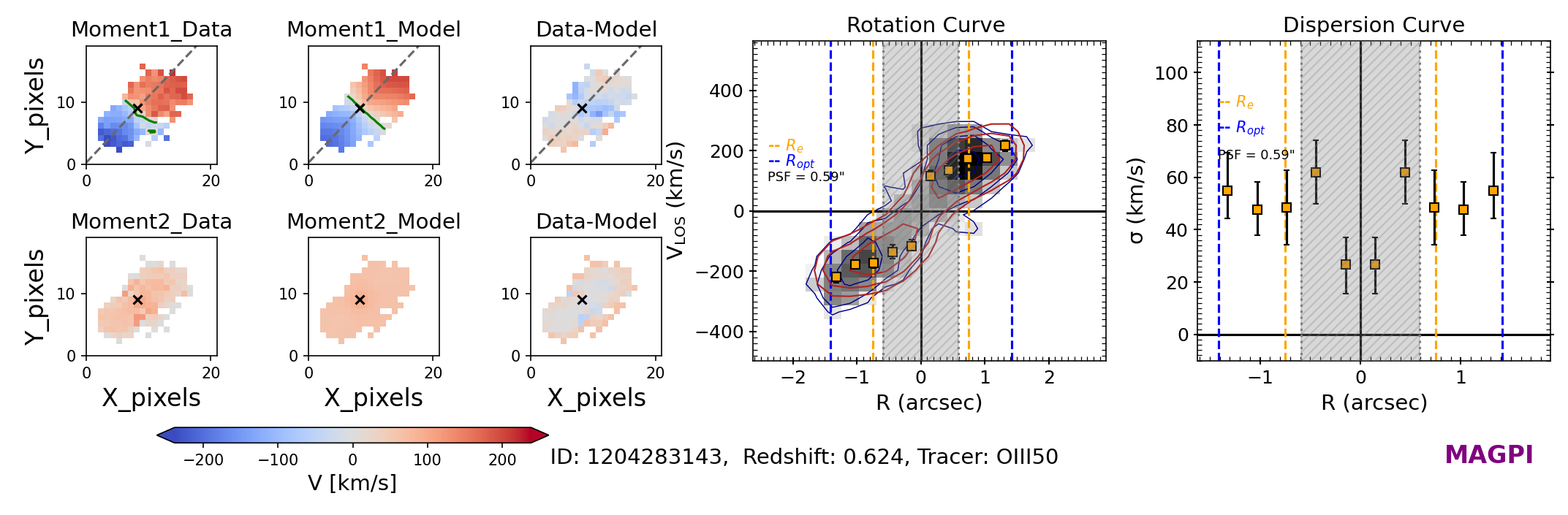} 
					
    \caption{Examples of kinematic modelling outputs. \textit{Row 1 \& 2, Columns 1--3:} moment-1 and moment-2 maps data, best-fit model, and residuals, respectively. The grey dashed line indicates the position angle, the black cross marks the central $(x, y)$ position, and the green line denotes the plane of rotation.  \textit{Column 4:} major axis position-velocity diagram, where the black shaded area with blue contours represents the data, red contours show the model, and orange squares with error bars indicate the best-fit line-of-sight rotation velocity derived by \texttt{3DBarolo}. The yellow and blue vertical dashed lines mark the effective radius ($R_{\rm e}$) and optical radius ($R_{\rm opt} = 1.89\,R_{\rm e}$), respectively. \textit{Column 5:} velocity dispersion profile as a function of radius. Error bars on the rotation and dispersion curves correspond to $1\sigma$ uncertainties. The PSF size is shown by gray hatched shaded region. Corresponding minor-axis PV diagrams demonstrating the accuracy of the recovered velocity dispersion profiles are available in the Supplementary Material.}
\label{fig:BBoutputs}
\end{center}
\end{figure*}
%%%%%%%%%%%%%%%%%%%%%%%%%%%%%%%%%%%%%%%%%%%%%
\subsection{Forward modelling of the data-cubes}
\label{sec:Barolo}
We conduct a comprehensive gas kinematics analysis of the MAGPI dataset using the 3D forward modelling approach implemented in 3DBarolo \citep{ETD15}.\footnote{We used 3DBarolo version 1.6.7, the latest version of 3DBarolo is publicly available at \href{https://editeodoro.github.io/Bbarolo/}{https://editeodoro.github.io/Bbarolo/}} The main advantages of modeling datacubes with 3DBarolo are that (1) it allows us to reconstruct the intrinsic kinematics for given initial guesses ring-by-ring, where each ring is generated in a 6D domain (three dimensions for the spatial location and three for the components of the velocity) and these rings are then projected into the datacube (i.e. $f(x,y,\lambda)$); (2) the 3D projected modeled datacubes are compared to the observed datacubes in 3D space; (3) it simultaneously incorporates instrumental and observational uncertainties (e.g., spectral smearing and beam smearing) in 3D space. This three-fold approach of deriving kinematics is designed to overcome the observational and instrumental biases faced in high‑$z$ observations/studies. In contrast, conventional 2D kinematic modelling derives velocity maps or rotation curves first and then attempts to correct for those biases a posteriori; this is less reliable for distant galaxies that have small angular sizes and moderate SNR \citep[for more details see][]{ETD16, GS21a}.

Following the methodology established in \citetalias{GS23}, we extract the gas kinematics using a custom optimization pipeline built around the \texttt{3DBarolo}. This pipeline systematically determines the optimal geometric parameters (kinematic center and position angle), systemic velocity, number of fitted rings, and emission line mask. The algorithm minimizes a custom loss function ($\ell$) designed to evaluate both the residual errors and the valid pixel coverage:
\begin{equation}
				\label{eq:loss_func}
				\ell  = \sqrt{  \overline{(X_D - X_M)^2}   } +  \frac{N[D]}{N[D: \neq 0, \neq \pm \infty]} + \frac{N[M]}{N[M: \neq 0, \neq \pm  \infty]} 
			\end{equation}
where $X_D$ and $X_M$ represent the data and model arrays, respectively. The first term calculates the root-mean-square error (RMSE) between the observation and the model. The second and third terms act as sparsity penalties: $N[\dots]$ denotes the total length of the array, while the denominators represent the number of valid (non-zero and finite) pixels. Consequently, models that generate excessively sparse or unphysical distributions are heavily penalized. To minimize $\ell$, we utilize the gradient-free \texttt{Nelder-Mead} algorithm available in the \texttt{scipy.optimize} library, which is highly effective for the non-linear, multi-modal objective functions inherent to kinematic modeling.			

Our pipeline constrains the parameters sequentially, explicitly tailoring the input data product to the parameter being optimized to maximize sensitivity. Specifically, the kinematic center ($x, y$) and the optimal emission mask (testing both \texttt{3DBarolo}'s internal \texttt{SEARCH} algorithm and a custom external segmentation mask) are constrained by minimizing the loss on the moment-0 map. The position angle (PA) is optimized using the moment-1 velocity field, while the systemic velocity ($V_{\rm sys}$) and the number of fitted rings (N$_{rings}$) are derived using the major-axis position-velocity (PV) diagram. The order of optimization is following: central position $\rightarrow$ mask $\rightarrow$ $V_{sys}$ $\rightarrow$ PA $\rightarrow$ N$_{rings}$. Once the loss function is minimized, the optimized parameter is fixed for next-step.

For each galaxy, the optimization explores a bounded parameter space anchored to the photometric priors. The kinematic centers are allowed to vary within $\pm 2$ pixels of the photometric center ($x_{\rm phot}, y_{\rm phot}$), and the position angle within $\pm 50^\circ$ of ${\rm PA}_{\rm phot}$. We evaluate models ranging from 2 to 20 radial rings. Finally, $V_{\rm sys}$ is iteratively evaluated within a $\pm 250\, \rm km\,s^{-1}$ window, simultaneously testing whether setting $V_{\rm sys}$ as a free or fixed parameter in the \texttt{3DFIT} task yields the lowest overall loss.

To ensure optimal spatial sampling within 3DBarolo, we set the radial distance between rings to half the instrumental beam size (\texttt{RADSEP = $FWHM_{_{PSF}}$/2}). Similarly, to account for the instrumental spectral broadening, we set the \texttt{LINEAR} parameter to the standard deviation of the Line Spread Function (\texttt{LINEAR = LSF/2.355}). During extraction, the \texttt{3DFIT} task utilizes an azimuthal normalization (\texttt{NORM = AZIM}), which forces the flux in the model rings to match the observed data, and a spatial weighting scheme (\texttt{WFUNC = 2}) that assigns higher statistical weight to pixels near the major axis. In the final fitting, the rotational velocity ($V_{rot}$) and velocity dispersion ($V_{disp}$) are always treated as free parameters. The systemic velocity ($V_{sys}$) is either fixed to the optimized grid-search value or left as a third free parameter, dictated by which specific configuration minimized the PV-diagram residuals during the optimization phase. The forthcoming public release of the MAGPI catalog will include all final derived parameters (including the list of free parameters and mask-type).

As demonstrated in \citetalias{GS23}, this approach effectively overcomes limitations of photometric geometrical parameters, which often poorly trace gas kinematics at higher redshifts due to morphological misalignments between stellar and gas components \citep[see also][]{W15,H17}. The detailed implementation, assumptions, and limitations of this technique are described in \citetalias{GS23} (Section 3.1 and Appendix A). In Figure~\ref{fig:BBoutputs}, we present representative examples of the kinematic modelling for MAGPI galaxies across various redshifts and emission line tracers. The left panels display the observed gas rotation velocity (moment 1) and velocity dispersion (moment 2) maps, alongside the corresponding best-fit models and residuals (data minus model). The residual maps are generally lie on and around zero, indicating an excellent agreement between the data and the models. Similarly, the central panels show that the position-velocity diagrams along the major axis exhibit a strong overlap between the observed data and the best-fit models.\footnote{To further validate the robustness of our kinematic models and the recovery of the velocity dispersion profiles, comprehensive minor-axis position-velocity (PV) diagrams for the full sample are provided in the Supplementary Material. Additionally, to ensure that no low-level systematic residuals are obscured by the broad kinematic color scale used in the main text, highly sensitive narrow-scale residual maps (e.g., $\pm 30$ km/s) are also included.} These results demonstrate the high quality of the kinematic fits and underscore the effectiveness of our modelling framework in recovering the intrinsic galaxy kinematics—including the radial velocity dispersion profiles illustrated in the right panels—while accurately accounting for beam smearing and other observational effects. We note that the  first point in the receding and approaching arms of rotation and dispersion curves are excluded from further analysis, as they fall within the region where 3DBarolo is unable to reliably model the kinematics due to seeing limit (for details see \cite{ETD15, ETD16}).

We note that in our sample, many galaxies exhibit asymmetric emission driven by data quality constraints or intrinsic morphological irregularities. Because our automated \textsc{3DBarolo} pipeline derives global kinematic models by fitting both the approaching and receding sides simultaneously, it is critical to ensure that the formal global velocity uncertainties accurately reflect the kinematics of both sides. To validate our global error budget, we performed independent kinematic fits for the approaching and receding arms of our entire sample. We then evaluated whether the formal global velocity uncertainty adequately captures the physical kinematic asymmetry between the two sides. The full results of this diagnostic are detailed in Appendix~\ref{sec:appendix_asymmetry}. To ensure statistical transparency, our final catalog includes a specific flag for highly asymmetric systems where the physical velocity difference between the approaching and receding arms exceeds the formal $2\sigma$ global uncertainty.

\subsection{Final sample}
\label{sec:Fsample}
To summarize the final sample selection, it is crucial to recall that the primary selection of galaxies for kinematic modelling was based on the following criteria: (1) confirmed spectroscopic redshift and detection of either (or all) of these emission lines:  $H\alpha$ ($6563\, \text{\AA}$), [OIII] ($4969\, \text{\AA}$), and [OIII] ($5008\, \text{\AA}$), (2) inclination angles within the range of $25^{\circ} \leq \theta_i \leq 75^{\circ}$, (3) integrated S/N $ > 3$, and (4) $log(M_{\rm star} \ [M_\odot]) = 8-11.5 $.
This primary selection yielded 434 galaxies that lie within $3\sigma$ range of \cite{speagle14} star-forming main sequence, as shown in Figure~\ref{fig:MSequence}, with the distribution of their physical properties illustrated in Figure~\ref{fig:MAGPI-hist}.

Following the kinematic modelling process (Section~\ref{sec:Barolo}), we applied secondary selection criteria  to remove galaxies with unreliable or insufficient kinematic constraints. Under this criteria, galaxies were excluded if they met the following conditions: (1) 3DBarolo+ optimization routine failed to converge (success = False) on a stable set of geometrical parameters within the maximum allowed iterations, resulting in unphysical fit; (2) No mask was created due to either a merger or low signal-to-noise per spaxels; (3) $\mathrm{R_{\rm max}< PSF}$, the spatial extent of the emission in these galaxies is insufficient relative to the beam size to resolve kinematic rings, precluding robust kinematic modeling; (4) $\mathrm{R_{\rm max}=PSF}$, in this case resulting kinematic models provide only two to three velocity measurements in \RCs, deemed insufficient for robust dynamical modelling. For further details and justification of the kinematic modelling quality control criteria, we refer the reader to \citetalias{GS23}. 

After applying the secondary selection criteria, we found that 18 galaxies failed to converge during the \textsc{3DBarolo} optimization, 9 galaxies lacked sufficient emission to construct a reliable mask, 6 galaxies exhibited $\mathrm{R_{\rm max} < PSF}$, and 4 galaxies had $\mathrm{R_{\rm max} = PSF}$. Consequently, the kinematically selected sample consists of 397 galaxies, as shown by the orange histograms in Figure~\ref{fig:MAGPI-hist}. Within this sample, 23 galaxies have kinematic measurements from all three emission lines, 1 galaxy from H$\alpha$ and [OIII]$\lambda4969$, 54 from H$\alpha$ and [OIII]$\lambda5008$, and 26 from [OIII]$\lambda4969$ and [OIII]$\lambda5008$. Additionally, 67, 6, and 89 galaxies are uniquely detected in H$\alpha$, [OIII]$\lambda4969$, and [OIII]$\lambda5008$, respectively. The final sample therefore includes 266 unique star-forming galaxies, hereafter (and in the plots) referred to as UIDs, which are highlighted with stars in Figure~\ref{fig:MSequence} and represented by the black hatched histograms in Figure~\ref{fig:MAGPI-hist}. For statistical robustness, however, all 397 kinematic measurements are included in the subsequent analysis, details are discussed in Section~\ref{sec:emline-RCs}.

%%%%%%%%%%%%%%%%%%%%%%%%%%%%%%%%%	
\subsection{Rotation curve robustness across emission lines}\label{sec:emline-RCs}
When comparing rotation curves derived from different emission lines, it is important to consider that each line may trace gas with different physical conditions and kinematic behaviors. Physically, these lines trace distinct phases of the ionized interstellar medium. The Balmer lines (H$\alpha$, H$\beta$) primarily trace the $T \sim 10^4$~K gas in star‑forming HII regions, where electron densities typically range from $n_e \sim 10^1$ to $10^3$~cm$^{-3}$ \citep{Osterbrock2006}. The [OIII] lines ($\lambda4960$, $\lambda5007$) trace a higher‑ionisation phase (requiring $\sim$35.1~eV), often corresponding to higher temperature gas ($T \approx 1.2-2 \times 10^4$~K) or excitation driven by non-thermal processes \citep{Osterbrock2006}. Consequently, these lines can potentially trace different kinematic behaviors, including non‑circular motions (e.g., outflows or turbulent broadening). For instance, in star‑forming disc galaxies, higher‑ionisation lines such as [OIII] may exhibit broader wings or blue‑shifted components driven by stellar or AGN feedback \citep[see e.g.,][]{Amorin2024}. If an emission line includes a significant outflow component, the derived rotation velocity and velocity dispersion may be biased, thereby affecting the subsequent mass decomposition and inferred DM fraction. In contrast, lower‑ionisation lines such as H$\alpha$ are often more closely associated with the rotationally supported disc plane and provide a cleaner tracer of the underlying gravitational potential \citep[e.g.,][]{Zurita2004}.

\begin{figure*}[h]
	\begin{center}
\includegraphics[angle=0,height=3.6truecm,width=5.0truecm]{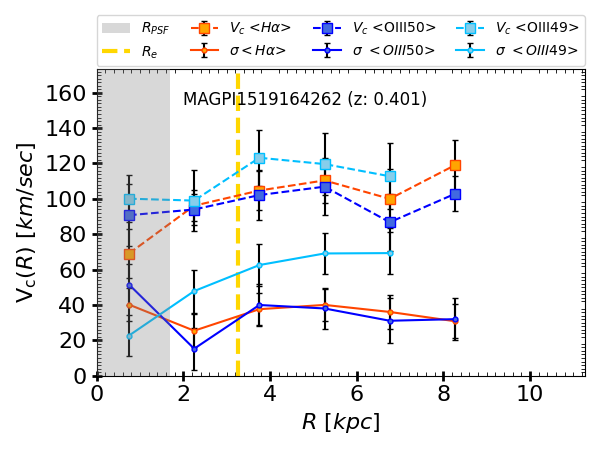} 
\includegraphics[angle=0,height=3.6truecm,width=5.0truecm]{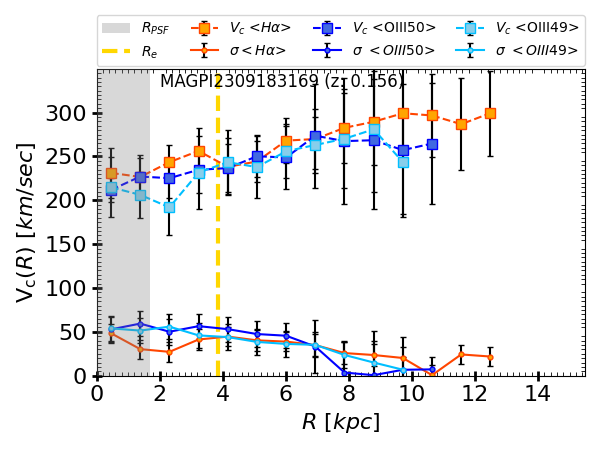} 
\includegraphics[angle=0,height=3.6truecm,width=5.0truecm]{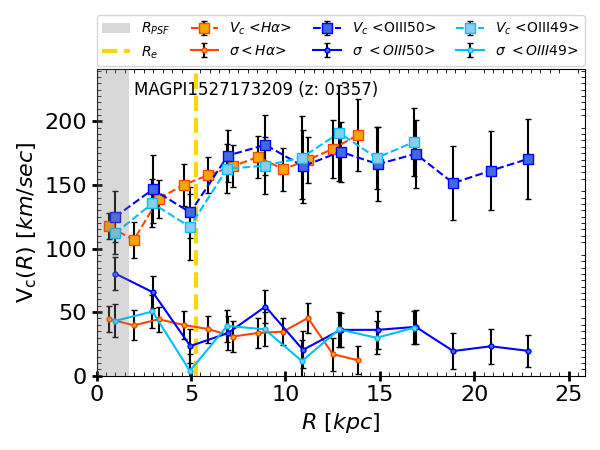} 

\includegraphics[angle=0,height=5.0truecm,width=12truecm]{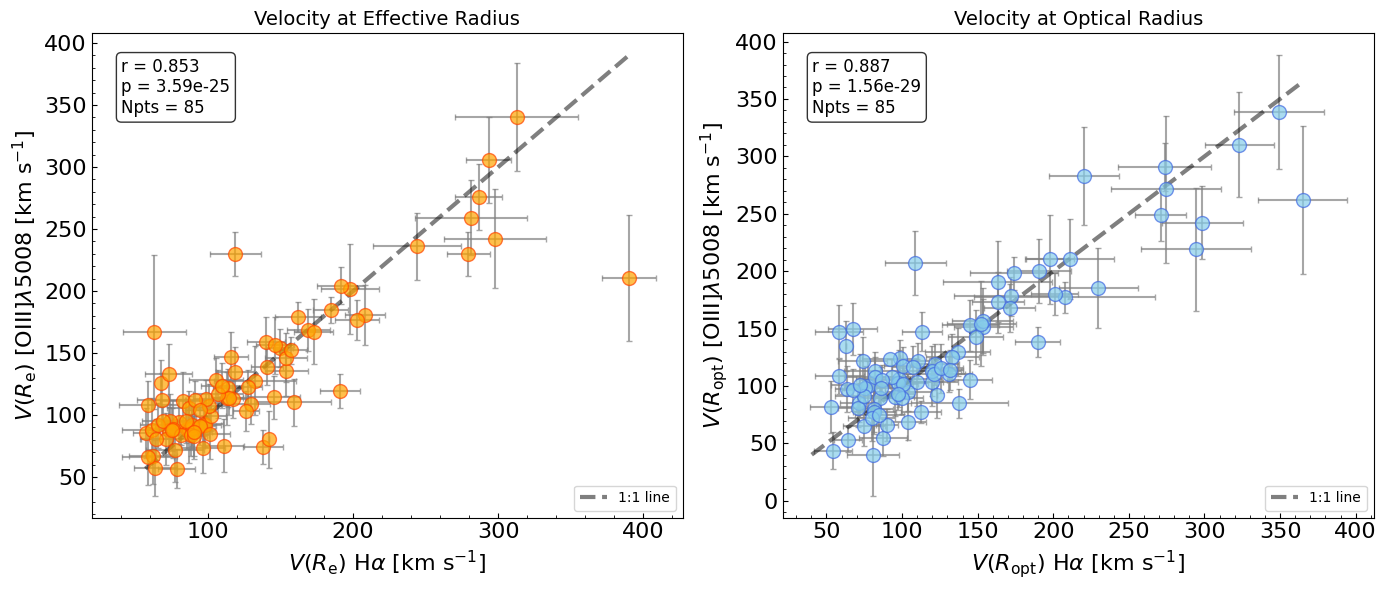} 

\includegraphics[angle=0,height=5.0truecm,width=7.0truecm]{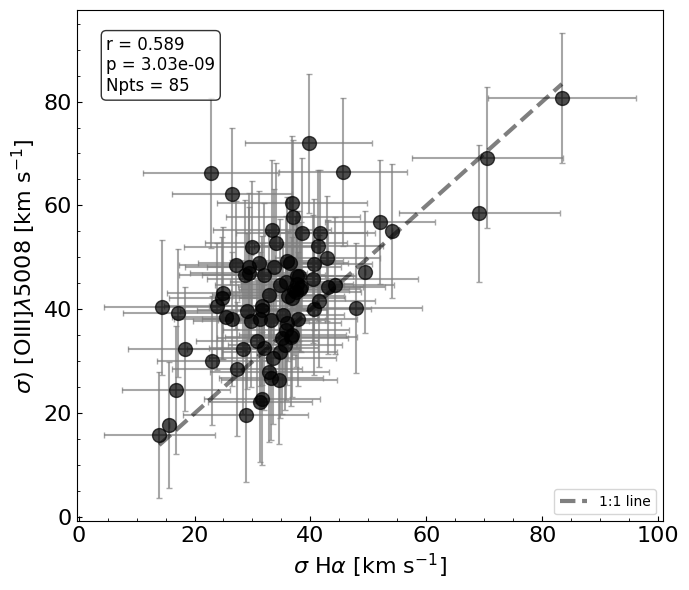} 

    \caption{{\em Row~1:} Example rotation curves derived from multiple emission lines. The circular velocity profiles are shown for H$\alpha$ (orange), [O~III]$\lambda$5008 (light blue), and [O~III]$\lambda$4960 (dark blue). The vertical yellow dashed line indicates the effective radius \( R_{\rm e} \), and the shaded gray region marks the beam-smearing-affected region \( R_{\rm PSF} \). {\em Row~2:} Comparison of velocities traced using H$\alpha$ and [O~III]$\lambda$5008 within effective radius and optical radius, left and right panels, respectively. {\em Row~3:} Comparison of observed velocity dispersion traced using H$\alpha$ and [O~III]$\lambda$5008.}
\label{fig:emline-RCs}
\end{center}
\end{figure*}

At the same time, the use of multiple emission lines, H\(\alpha\), [OIII]\(\lambda5008\), and [OIII]\(\lambda4960\), provides key advantages. First, it allows us to build a larger and a broad redshift range sample. Especially, when working at high-redshift where H\(\alpha\) lie outside the MUSE spectral range, [OIII] remains accessible; hence, become essential for tracing ionized gas kinematics at high-$z$. Second, the spatial extent of the emission varies between lines: [OIII]\(\lambda4960\) is typically detected in the more central regions compared to H\(\alpha\) and [OIII]\(\lambda5008\), thus providing complementary constraints on the inner rotation curve. Our final sample includes 23 galaxies for which all three emission lines are available, and 81 galaxies with either of the two combinations of emission lines as mentioned in Section~\ref{sec:Fsample}.

%original version
%As demonstrated in Figure~\ref{fig:emline-RCs}, the subset of galaxies with all three emission lines show that the rotation curves overlap within uncertainties and that no systematic residuals remain. Moreover, their circular velocities within $R_e$ and $R_{\rm opt}$ are consistent within 0.05 dex. That is, these different emission lines, despite variations in line strength and ionization potential, consistently trace the same underlying gravitational potential ($\propto V_c^2$). This agreement indicates that any (or all) of these lines can be reliably used to derive gas kinematics and the DM fractions at all redshifts probed in this work. Nonetheless, we caution that differential line‑broadening or hidden non‑circular motions could still introduce additional systematic uncertainties. We verify consistency for the subset of galaxies with both H\(\alpha\) and [OIII]\(\lambda5008\) by comparing their rotation curves   and their circular velocities at different radii, yet fully quantifying any residual bias remains challenging without higher spectral resolution and advanced kinematic modelling tools, as highlighted in Section~\ref{sec:errors}. 

As demonstrated in Figure~\ref{fig:emline-RCs}, the subset of galaxies with all three emission lines show that the rotation curves overlap within $1\sigma$ uncertainties and that no systematic residuals remain. Moreover, their circular velocities within $R_{\rm e}$ and $R_{\rm opt}$ are consistent within 0.05 dex. That is, these different emission lines, despite variations in line strength and ionization potential, consistently trace the same underlying gravitational potential ($\propto V_c^2$). Similarly, the intrinsic velocity dispersion ($\sigma$) measurements are broadly consistent across the extended disk regardless of the chosen tracer. However, we note that the [\text{O~III}] $\lambda5007$ and [OIII]\(\lambda4960\) lines occasionally exhibits slightly elevated velocity dispersions (by $\sim 10-30 \text{ km/s}$) compared to H$\alpha$. Importantly, this discrepancy does not impact the derived circular velocity profiles, as it is naturally accounted for by our gas pressure gradient corrections (see Figure~\ref{fig:emline-RCs}, first \RC\ example).  This agreement in the \RCs\ indicates that any (or all) of these lines can be reliably used to derive gas kinematics and the DM fractions at all redshifts probed in this work.  Nonetheless, we caution that differential line‑broadening or hidden non‑circular motions could still introduce additional systematic uncertainties. We verify consistency for the subset of galaxies with both H$\alpha$ and [\text{OIII}]$\lambda5008$ by comparing their rotation curves, velocity dispersion profiles, and their circular velocities at different radii, yet fully quantifying any residual bias remains challenging without higher spectral resolution and advanced kinematic modelling tools, as highlighted in Section~\ref{sec:errors}.

Finally, to prevent multiple occurrences of the same galaxy in the statistical representation of the dark matter fraction trends, we assign a single representative value to galaxies with measurements from more than one emission line. For such galaxies, the dark matter fraction is computed using an inverse-variance (error-weighted) average:
\begin{equation}
x_i^{\mathrm{rep}} =
\frac{
\sum_{j=1}^{N} x_i^{(j)} \, / \, \bigl(\sigma_i^{(j)}\bigr)^2
}{
\sum_{j=1}^{N} 1 \, / \, \bigl(\sigma_i^{(j)}\bigr)^2
},
\label{eq:weighted_mean}
\end{equation}
where \(x_i^{(j)}\) denotes the dark matter fraction of the \(i\)-th galaxy measured from the \(j\)-th emission line, \(\sigma_i^{(j)}\) is the corresponding measurement uncertainty, and \(N\) is the number of available emission-line measurements for that galaxy. The propagated uncertainty on the representative value is given by

\begin{equation}
\sigma_i^{\mathrm{rep}} = \frac{1}{\sqrt{\sum_{j=1}^{N} \frac{1}{(\sigma_i^{(j)})^2}}}
\label{eq:weighted_error}
\end{equation}

This procedure ensures that each galaxy contributes a single statistically independent data point to the population-level analysis while optimally incorporating information from multiple emission-line tracers.

%%%%%%%%%%%%%%%%%%%%%%%%%%%%%%%%%%%%%%%%%%
\section{Method}
\label{sec:method}
In this section, we outline the methodology used to derive DM fractions in our galaxy sample.% Firstly, we perform a mass decomposition method for the stellar disk and bulge, as well as gas disk components. Then, we introduce the formalism used to estimate the DM fraction.

\subsection{Rotation curve modelling \& DM fraction method}
\label{sec:MModel}
To disentangle the baryonic components and estimate the DM fraction from the \RCs, we first applied the pressure gradient correction, a method introduced in \cite{GS21a} and briefly described in Appendix~\ref{sec:PGC}. The pressure gradient corrected \RCs\ provide the intrinsic total circular velocity profile, $V_c(R)$, of the system. In a self-consistent dynamical model, this total circular velocity is composed of the contributions from baryons (stars and gas) and DM, summed in quadrature to account for their distinct gravitational potentials:

\begin{equation} 
\label{eq:vcQuadrature}
V_c^2(R) = V_{\rm disk}^2(R) + V_{\rm bulge}^2(R) + V_{\rm gas}^2(R) + V_{\rm DM}^2(R),
\end{equation}

where the baryonic velocity contribution, $V_{\rm bar}^2(R)$, is the sum of the stellar and gaseous components:

\begin{equation}
\label{eq:vbarQuadrature}
V_{\rm bar}^2(R) = V_{\rm disk}^2(R) + V_{\rm bulge}^2(R) + 1.33 \ V_{\rm gas, HI}^2(R) + V_{\rm gas, H_2}^2(R).
\end{equation}

where, $V_{\rm disk}(R)$ and $V_{\rm gas}(R)$ are derived using a flattened thin-disk potential (e.g., \cite{Freeman}), while $V_{\rm bulge}(R)$ and $V_{\rm DM}(R)$ are calculated assuming spherical symmetry ($V_i^2 = G M_i/R$). In these formulations, the atomic gas contribution ($V_{\rm gas, HI}$) includes a factor of 1.33 to account for the Helium fraction. For the molecular gas ($V_{\rm gas, H_2}$), the Helium contribution is already incorporated into the scaling relations of \cite{Tacconi2018}, as discussed in Appendix~\ref{sec:Mbar}. Under this formalism, the DM fraction at a given radius $R$ is defined as the ratio of the squared circular velocity of the dark matter halo to the total squared circular velocity:

\begin{equation}
\label{eq:Fdm}
f_{{\rm DM}} (<R) = \frac{V_{_{\rm DM}}^2(R)}{V_c^2(R)} = 1 - \frac{V_{\rm bar}^2(R)}{V_c^2(R)}
\end{equation}

In the following sub-sections, we present an approach to model the aforementioned baryonic velocity components from total circular velocity profile (i.e., pressure gradient corrected \RC).

%This formulation allows us to compute the DM mass distribution (and fraction) by removing the contribution of baryons from the total velocity. In the following sections, we present an approach to model the aforementioned components from pressure support corrected \RCs. We note that the circular velocity measured at $R_{\rm opt}$—where the rotation curve is typically flat—is adopted as the characteristic circular velocity ($V_c$) of a galaxy. 

%This velocity-based decomposition allows us to compute the DM contribution as the residual of the total dynamical support:\begin{equation}V_{\rm DM}^2(R) = V_c^2(R) - V_{\rm bar}^2(R).\end{equation}Finally, the DM fraction within a given radius $R$ is self-consistently defined as the ratio of the squared velocities:\begin{equation}f_{_{\rm DM}}(R) = \frac{V_{\rm DM}^2(R)}{V_c^2(R)}.\label{eq:fdm_quad}\end{equation}This formulation ensures that the geometric effects of the flattened disk components are properly accounted for, providing a more accurate estimate of the dark matter contribution compared to a purely spherical mass assumption. We adopt the circular velocity measured at the optical radius ($R_{\rm opt}$)—where the rotation curve is typically flat—as the characteristic circular velocity ($V_c$) of the galaxy.

\paragraph{\textbf{Stellar and gas disk:}} As demonstrated in Section \ref{sec:data}, the MAGPI sub-sample analysed here lies on and around the the star-forming MS. While there is some structural diversity on the MS, the majority of galaxies have been shown to be well described by discs \citep[e.g.,][]{Wuyts2011b,Forster2020}. Moreover, as shown in Figure~\ref{fig:PGS_plot}, the galaxies in our sample are rotationally supported systems, i.e., consistent with late-type spirals on the main sequence that host well-ordered stellar disks \citep[e.g., see][]{Gurovich2010}. Therefore, in modelling the disk component, we assume that stars and gas follow an exponential  profile, as described by the Freeman disk \citep{Freeman}:

\begin{equation}
	\label{eq:Mcomp}
	V_{\_}^2(R)= \frac{1}{2} \Big( \frac{GM_{\_}}{R_{\rm scale}} \Big) \  \Big( x \Big)^2 \  [I_0K_0 - I_1K_1],
\end{equation}

%	V_{\_}^2(R)= \frac{1}{2} \Big( \frac{GM_{\_}}{R_{\rm scale}} \Big) \  \Big( \alpha \frac{R}{R_{\rm scale}} \Big)^2 \  [I_0K_0 - I_1K_1],
where $M_{\_}$ and $R_{\rm scale}$ are the total mass and the scale length of the different components (stars, H2, and HI), respectively, $x=\frac{R}{R_{\rm scale}}$, $I_n$ and $K_n$ are modified Bessel functions computed at $x=1.6$ for stars and H2-gas, and $x = 0.53$ for HI-gas \citep[c.f.][]{PS1996}.\footnote{Unlike the highly extended neutral hydrogen (HI) disk, molecular gas (H$_2$) typically follows a surface density profile that tightly traces the stellar disk \citep[e.g.,][]{Leroy2008, Schruba2011}. Assuming comparable exponential scale lengths at intermediate redshifts, we adopt the same Bessel function argument for the H$_2$ kinematic component as derived for the stellar disk.} For example, in Equation~\ref{eq:Mcomp} if we work with stellar disk component $R_{scale}=R_{D}$ and $M^{\rm tot}_{\_} = M^{\rm tot}_{disk} = M^{\rm tot}_{\rm star} - M_{\rm bulge}$ (likewise for other components). We note that the total stellar mass is estimated using SED fitting, while atomic and molecular gas masses are derived from scaling relations (see Section~\ref{sec:data} \& Appendix~\ref{sec:Mbar}).

The stars are distributed within the stellar disk scale length, $R_{\rm D} = 0.59\times R_{\rm e}$, determined from photometry (Section~\ref{sec:data}). Regarding the gaseous components, the lack of resolved multi-phase imaging at intermediate redshifts necessitates analytical approximations for their spatial distribution. We assume the molecular gas ($H_2$) traces the star-forming ionized gas reservoir, setting $R_{\rm H2} \approx R_{\rm gas}$, where $R_{\rm gas}$ is determined by fitting the $H_\alpha$ surface brightness \citep[][see Appendix-C]{GS21b}. For the atomic gas ($HI$), we adopt a radial scale length of $R_{\rm HI} \approx 2 \times R_{\rm H2}$. This scaling is physically motivated by the high average gas masses of our sample ($\langle \log(M_{HI}/M_{\odot}) \rangle \approx 9.6$); for systems in this mass range, an $R_{HI}/R_{H2}$ ratio of approximately 2 is consistent with empirical angular momentum-mass relations \citep[e.g.,][]{ManceraPina2021, Geesink2025}. While these represent approximations, they provide a robust and self-consistent framework for estimating the gaseous contribution to the total dynamical support.

%We note that the total stellar mass is estimated using SED fitting and the total atomic and molecular gas masses are estimated using scaling relations, for details see Section~\ref{sec:data} \& Appendix~\ref{sec:Mbar}, respectively. The stars are assumed to be distributed within the stellar disk radius (i.e., in this case $R_{scale} = R_{\rm D} = 0.59\times R_{\rm e}$) known from photometry, discussed in Section~\ref{sec:data}. The molecular gas is generally distributed outward through the stellar disk (up to the length of the ionized gas $R_{\rm gas}$); therefore, we take $R_{\rm H2} \equiv R_{\rm gas}$. We estimate the gas scale length $R_{\rm gas}$ by fitting the $H_\alpha$ surface brightness, see \cite[][see Appendix-C]{GS21b}.  Moreover, studies of local disk galaxies have shown that the surface brightness of the HI disk is much more extended than that of the $H_2$ disk \citep[][see their Fig. 5]{Jian2010}; see also \cite{Leroy2008} and \cite{Cormier2016}. Therefore, we assume $R_{\rm HI}=2 \times R_{\rm H2}$, which is a rough estimate, but still reasonable, considering that at high-$z$ very little information is available on the $M_{\rm HI }$ (or $M_{\rm H2}$) surface brightness distribution.

\begin{figure*}
	\begin{center}
\includegraphics[angle=0,height=5.5truecm,width=7.5truecm]{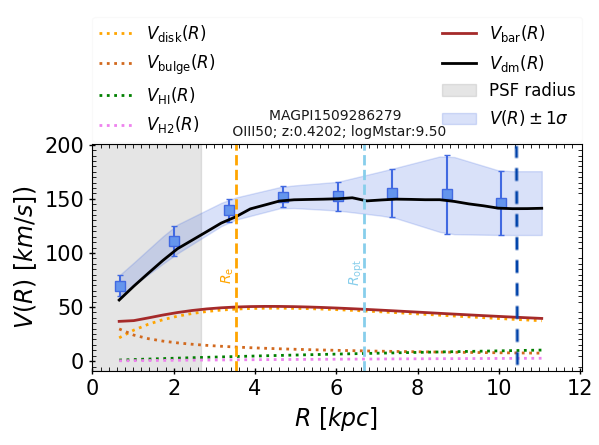} 
\includegraphics[angle=0,height=5.5truecm,width=7.5truecm]{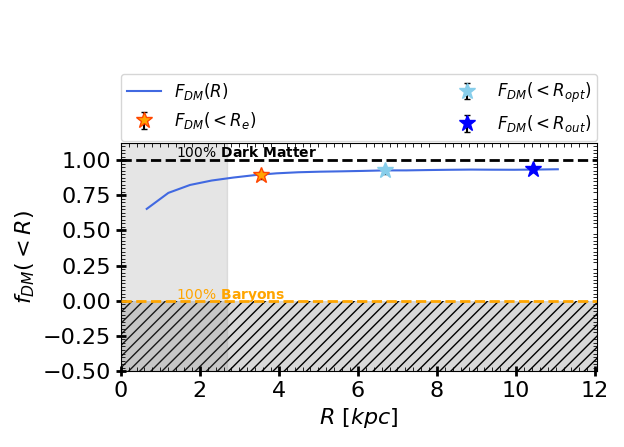}
\caption{%Kinematic and mass decomposition properties of the MAGPI galaxy sample. %{\em Top panel:} Stellar Tully-Fisher relation for 397 galaxies in the sample, showing a best-fit slope of 3.9 and an intrinsic scatter of 0.29\,dex. The dashed black line indicates the best-fit relation, and the blue points represent individual galaxies with associated uncertainties. 
Mass modeling (velocity decomposition) example of the MAGPI galaxy sample. 
{\em Left panel:} rotation velocity profile for MAGPI1509286279 (\( z = 0.4202 \), \(\log(M_{star}\, M_\odot) = 9.50\)). The gas pressure gradient corrected rotation velocity  (blue squares), total baryonic mass velocity  (red line), stellar disk velocity (orange line), bulge velocity (brown line), atomic gas velocity (green dotted line), molecular gas gas (pink dotted line), and DM velocity (black line) are shown. The vertical gray shaded area and dashed lines in orange and sky blue, and blue indicate the PSF, effective radius (\( R_{\rm e} \)), optical radius (\( R_{\rm opt} \)), and outer radius (\( R_{\rm out} \)), respectively. {\em Right panel:} Corresponding DM fraction as a function of radius. The solid blue line shows the cumulative DM fraction \( f_{_{\rm DM}}(<R) \), with the horizontal black and orange dashed lines indicating 100\% DM (top) and 100\% baryons (bottom), respectively. Stars indicate the DM fraction within the \( R_{\rm e} \) (orange), \( R_{\rm opt} \) (cyan), and \( R_{\rm out} \) (blue).}
\label{fig:STFR-Mdyn-Fdm}
\end{center}
\end{figure*}

\paragraph{\textbf{Stellar bulge:}} 
Accurately modelling the stellar bulge component is essential for galaxies at intermediate redshifts, as many star-forming galaxies exhibit bulge signatures that can substantially affect the inner mass distribution and gravitational potential \citep{Dimauro2018, Morselli2017,Trevor2014}. Neglecting the bulge contribution may lead to overestimation of the DM fraction in the central regions. Therefore, we provide first order approximation to model the bulge mass distribution. A detailed bulge-disk decomposition will be presented in a follow up work of Mauro et al. and Thater et al. (in preparation). In this work, we use an analytic Hernquist profile \citep{Hernquist1990}, which provides a realistic description of spherical bulges, in which the enclosed mass as a function of radius is given by:
\begin{equation}
	\label{eq:Mbulge}
	\begin{array}{r@{}l}
M_{\rm bulge}(r) & = M_{\rm bulge} \frac{R^2}{(R + R_{bulge})^2},\\ \\
V^2_{\rm bulge} & = \frac{GM_{\rm bulge}}{R},
\end{array}
\end{equation}
where \( M_{\rm bulge} \) is the total bulge mass and \( R_{bulge} \) is the bulge scale radius. The total bulge mass is estimated using bulge-to-total stellar mass ratio of star-forming main-sequence galaxies given by \citep{Morselli2017}, which has an average scatter of less than $0.1$ dex. The scale radius is computed following the empirical relation from \cite{Hon2023}, which links bulge mass to its effective radius as:
\begin{equation}
	\label{eq:Rbulge}
R_{\rm bulge} = 10^{0.84 \log_{10}(M_{\rm bulge}) - 8.81},
\end{equation}
where \( R_{\rm bulge} \) is in kiloparsecs and \( M_{\rm bulge} \) in solar masses. The relation has an intrinsic scatter of 0.22 dex in the $R_{\rm bulge}$ direction. While this relation was calibrated for local bulges, it remains a reasonable approximation for \sfgs\ at \( 0 < z \lesssim 1.0 \), as studies have shown that the structural properties of bulges, including size–mass relations, evolve only mildly at these redshifts \citep[e.g.,][Fig.~11]{Dimauro2019}. Therefore, applying this relation provides a physically motivated, first-order, estimate of the bulge scale radius in the absence of direct bulge size measurements at intermediate-redshifts. %Incorporating the bulge mass in our models enables a more accurate decomposition of baryonic and DM contributions, particularly within the inner regions where the bulge dominates the gravitational potential.
\\
\\
In Figure \ref{fig:STFR-Mdyn-Fdm} (left panel), we illustrate this decomposition for a representative galaxy  MAGPI 1509286279 (\( z = 0.4202 \), \(\log(M_{star}\, M_\odot) = 9.50\)). Its rotation curve (pressure corrected) is plotted alongside the individual velocity contributions from the stellar disk, bulge, gas disk, and dark matter components.  The right panel of Figure \ref{fig:STFR-Mdyn-Fdm} shows the cumulative DM fraction as a function of radius. We observe a monotonic increase in $f_{_{\rm DM}}$, marking the transition from a baryon-influenced center to a halo-dominated regime in the outskirts. Our estimates of the DM fraction are calculated up to the optical radius ($R_{\rm opt} = 1.89 \times R_{\rm e}$). While we initially aimed for a broader radial coverage, the data for the outermost regions ($3 \times R_e$) is not uniformly available across the entire MAGPI sample. For the purpose of our analysis, we define the region within $R_{\rm e}$ as the \textit{inner region} and the region between $R_{\rm e}$ and $R_{\rm opt}$ as the \textit{outskirts}. The total circular velocity measured at $R_{\rm opt}$ is denoted as characteristic circular velocity ($V_c$).

\subsection{Uncertainties \& limitations} 
\label{sec:errors}
Uncertainties on the velocity measurements are provided by \textsc{3DBarolo} using Monte Carlo sampling. Morphological parameters (e.g., the effective radius), stellar masses and star-formation rates incorporate statistical uncertainties derived via Monte Carlo sampling. The statistical uncertainties on bulge mass and bulge scale length are fixed to 0.01 dex and 0.22 dex, respectively. For each galaxy, the fraction of DM is computed by propagating all relevant uncertainties via a Monte Carlo sampling with 10000 realizations. In each realization the input values (velocity, stellar \& gas mass, bulge mass, SFR, and scale lengths ) are randomly drawn from their respective error distributions, and systematic uncertainties (such as those on SFR estimates $\sim 0.1$ dex) and the intrinsic scatter of gas-mass scaling relations are included ($\sim 0.3$ dex for both HI and H$_2$, independently). Thus the derived distribution of dark-matter fraction encapsulates both statistical and systematic offsets. To derive the average dark-matter fraction of the sample, we compute both the weighted mean and the median. Uncertainties on these average values are estimated via bootstrap resampling and are reported as the 68\% confidence intervals. 

We emphasize that the study of DM fraction at \hz is very challenging and requires significant refinements in both baryonic constraints and kinematic modeling. For example, SED fitting techniques, which are commonly used to estimate stellar masses and SFRs, exhibit substantial offsets from each other \citep{Leja2019, Pacifici2023}. These discrepancies not only lead to larger uncertainties in stellar masses and SFRs, but also impact the gas mass scaling relations. For instance, \cite{Tacconi2018}, molecular gas mass scaling relation, uses the \cite{speagle14} main-sequence relation ($M_*$-SFR plane). At \(z \sim 0.1\), the \cite{speagle14} main sequence differs from \cite{Chang2015} by \(\sim 0.4\) dex, and at \(z \sim 1\), it is \(\sim 0.25\) dex off from the 3DHST study \citep{Nelson2021}. 
%At higher redshifts, this offset is not yet quantified. 
Therefore, the systematic offset and observed scatter in gas scaling relation, which relay on main-sequence relation is complex to accurately quantify, and beyond of the scope of this study.

A limitation of this work is that, from a dynamical perspective, the rotationally supported thin disk is treated as an exponential ($n=1$) component, while the excess central light is modeled as a pressure-supported Hernquist bulge. Nevertheless, we confirm the structural validity of this decomposition across our entire sample in Appendix~\ref{sec:Mbulge-Mdisk}, showing a tight one-to-one correlation with the global photometry at both $R_{\rm e}$ and $R_{\rm opt}$. We explicitly opt for this physically constrained setup to avoid the severe parameter degeneracies (such as the unphysical trading of mass between the bulge and inner disk) that inherently limit unconstrained multi-component profile fitting/optimization when applied to seeing-limited data.

We adopt $R_{opt} = 1.89 R_e$ as the characteristic radius for calculating total circular velocities and dark matter fractions. While studies of local low-mass galaxies often utilize larger radii ($3-5 R_e$) to reach the flat part of the rotation curve \citep[e.g.,][]{Romanowsky2012,ManceraPina2026}, such extended mapping is restricted by the surface brightness limits of MUSE at intermediate redshifts. Moreover, because the scaling relations (radii transformation, $R_e$ to $R_{opt}$) between morphological and kinematic radii are not yet calibrated at these intermediate redshifts. Therefore, introducing a mass-dependent reference radius would introduce a new source of systematic uncertainty into our analysis, as the relationship between $R_e$ and the transition to the `flat' part of the rotation curve may evolve over cosmic time. However, by adopting a uniform definition of $R_{opt}$ for the entire sample, we ensure a cleaner statistical and radii based comparison in the context of galaxy-evolution. We also note that since the dark matter halo is cumulative $f_{_{\rm DM}}$ estimates at larger radii would expectedly yield higher fractions, but would be less robust across our full sample due to decreasing SNR and extrapolation.

			Due to the lack of direct measurements of HI gas at high redshifts (\(z > 1\)), accurate gas scale length and masses at \(z > 1\) are still unavailable. At \(z \leq 1\), HI mass scaling relations are derived using stacking analyses, as current observing facilities only allow HI signal detection without resolution \citep{Chowdhury2022}. Hence, it gives only global gas mass but scale length is still impossible to measure. Moreover, these HI scaling relations are general and not specifically for star-forming `disk-like' galaxies. That is, we can use them as a first order approximation, but they are not accurate.  Refining HI mass and scale length will requires future observations from the Square Kilometer Array \citep[e.g.,][and references therein]{SKA2017,SKA2022}.
			
			Moreover, current kinematic modeling techniques are limited by their inability to fully account for non-circular motions, gas inflows, and outflows \citep{Oman2019}. As a result, achieving accurate estimates of the total dynamical mass at given radius remains challenging. Additionally, the stellar bulge is typically unresolved; thus, its mass and scale length are inferred from empirical scaling relations \citep{Morselli2017, Hon2023}, each introducing their own intrinsic scatter.

Another caveat to our dynamical modeling is that the spatial sampling of our IFU data inherently oversamples the observational beam. As highlighted by recent literature \citep[e.g.,][]{Posti2022, Chase2026}, this oversampling introduces spatial covariance between adjacent pixels. Because our automated fitting pipeline treats adjacent spaxels as independent and does not explicitly calculate the spatial covariance matrix, the purely statistical uncertainties derived from the fitting routine may be underestimated. While we attempt to mitigate this by empirically validating our global error budget against the macroscopic kinematic asymmetries of the disks (see Appendix~\ref{sec:appendix_asymmetry}), the lack of explicit pixel covariance modeling remains a systematic caveat in our formal uncertainty budget.
			
Addressing aforementioned issues will require very high-resolution large galactic-scale surveys at high redshifts, which are currently not feasible even in optical and near-infrared astronomy. Further progress might only be possible with the Extremely Large Telescope and Square Kilometer Array. Nevertheless, given the best possible information available on  baryonic content, observed kinematics, and their uncertainties, we present conservative DM fraction estimates (and their uncertainties) to the best of our abilities. However, we acknowledge that these findings require further refinement, which will be achieved as the field progresses.

%%%%%%%%%%%%%%%%%%%%%%%%%%%%%%%%%%%%%%%%%%
\section{Results \& discussion}
\label{sec:result}
In this section, we examine the relationship between the dark matter, baryonic content, and halo assembly properties. We also present the DM evolution with cosmic time. Specifically, we analyze the scaling relations of the DM fraction measured within the effective radius ($R_e$, refer to as inner region) and the optical radius ($R_{\rm opt}$, refer to as outskirts). We include only galaxies for which the radius of interest ($R_e$ or $R_{\rm opt}$) exceeds the PSF FWHM, to ensure reliable measurements. As a result, DM fraction within $R_e$ and $R_{\rm opt}$ is presented for 181 and 246 galaxies, respectively.% out of the total sample of 266. 

\subsection{Dark matter dynamics: scaling $f_{_{\rm DM}}$ with $V_c$}
\label{sec:Fdm-Vc-Mstar}
\begin{figure*}[h]
	\begin{center}
\includegraphics[angle=0,height=5.0truecm,width=14.0truecm]{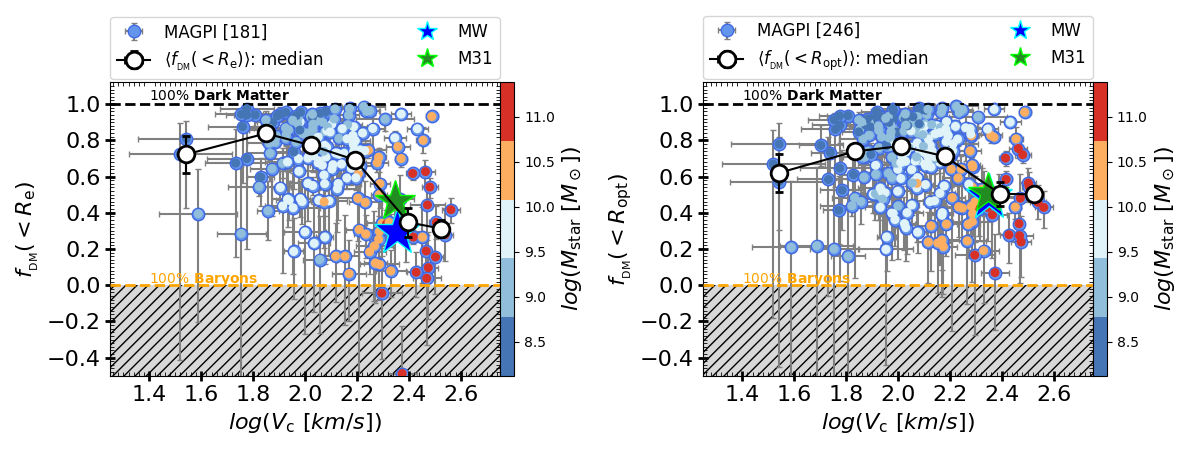} 
\caption{DM fraction (\( f_{_{\rm DM}} \)) as a function of circular velocity (\( V_{\rm c} \)). The left and right panels shows the DM fraction within the effective radius (\( R_{\rm e} \)) and optical radius (\( R_{\rm opt} \)), respectively. 
In both panels, individual galaxies are color-coded by stellar mass (\( M_{star} \)). The white circles with error bars represent the median \( \langle f_{_{\rm DM}} \rangle \) values in bins of \( V_{\rm c} \). The errors on the datasets (individual and average) are 68\% confidence intervals. The positions of the Milky Way (MW; blue star, \cite{MilkyWay}) and Andromeda (M31; green star, \cite{Andromeda}) are shown for comparison with local massive disk galaxies. Horizontal black and orange dashed lines indicate the 100\% DM and 100\% baryons limits, respectively. The gray hatched region corresponds to unphysical negative DM fractions. These panels show that the majority of galaxies are DM dominated and illustrate the dependence of the DM content on galaxy mass and rotation velocity, with a clear increase in \( f_{_{\rm DM}} \) at lower $M_{star}$ and $V_c$.
%In both panels, the numbers 274 and 397 indicate the total number of galaxies plotted that have scale radii larger than the PSF FWHM.
 }
\label{fig:Fdm-Vc-Mstar}
\end{center}
\end{figure*}

\begin{figure*}[h]
	\begin{center}
\includegraphics[angle=0,height=6.0truecm,width=9.5truecm]{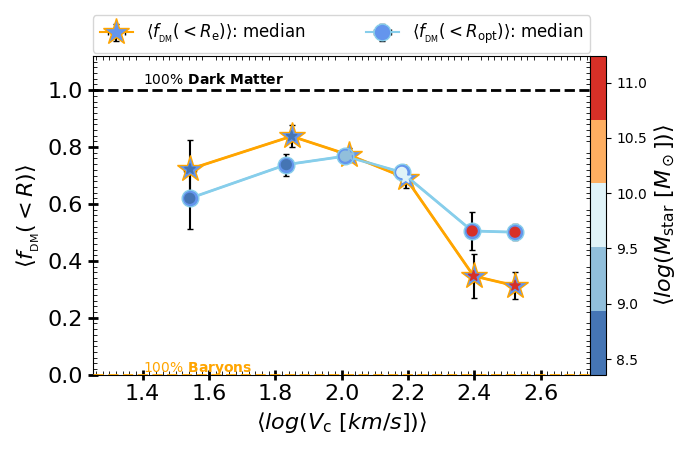} 
\caption{Averaged DM fraction (\( \langle f_{_{\rm DM}} \rangle \)) within the effective radius (stars) and optical radius (circles), computed in bins of circular velocity (\( V_{\rm c} \)). Each \( \langle f_{_{\rm DM}} \rangle \) value is color-coded for median stellar mass obtained in the corresponding velocity bin. These results demonstrate that, in high-velocity systems compared to low-rotation-supporters, the DM fraction within the effective radius is reduced by approximately 30-40\% (including errors) relative to its value at the optical radius, whereas the reduction within the optical radius, for such systems, is limited to about 10\%.}
\label{fig:Fdm-Vc-Mstar2}
\end{center}
\end{figure*}

In Figure~\ref{fig:Fdm-Vc-Mstar}, we present the relationship between the DM fraction and the circular velocity of galaxies. The DM fraction is computed within \( R_{\rm e} \) (inner region; left panel) and \( R_{\rm opt} \) (outskirts; right panel). To provide a reference for the baryonic component, galaxies are color-coded by their stellar masses. The error bars reflect the full propagation of mass-model uncertainties at the  $1\sigma$ level. We notice that the low-mass galaxies (\( M_{\rm star} < 10^{9.5}\, M_\odot \)) are typically low-velocity systems (\( V_{\rm c} < 100\, \mathrm{km\,s^{-1}} \)) exhibit a median DM fraction of \( \langle f_{_{\rm DM}}(<R_{\rm e}) \rangle = 0.77 \pm 0.1 \) and \( \langle f_{_{\rm DM}}(<R_{\rm opt}) \rangle = 0.73 \pm 0.05 \). That is, at low $V_c$ and low stellar masses, galaxies are DM dominated as expected for low-mass systems where baryons contribute less to the overall potential. In contrast, high-velocity systems (\( V_{\rm c} > 100\, \mathrm{km\,s^{-1}} \)) are predominantly associated with intermediate- to high-mass galaxies (\( 10^{9.5} \leq M_{\rm star}\, [M_\odot] \leq 10^{11.5} \)). In these systems, the DM fraction within \( R_{\rm e} \) decreases significantly with increasing stellar mass, while the decline in the outskirts (\( R_{\rm opt} \)) is modest. We show this quantitatively in Figure~\ref{fig:Fdm-Vc-Mstar2}, where we notice that the DM fraction for high-velocity systems is reduced by approximately 55\% in the inner regions, compared to only about a 30\% reduction in the outskirts (see the blue median-fit compared to orange). This relatively large change in \( f_{_{\rm DM}} \) in the inner regions of galaxies likely reflects the influence of baryonic feedback processes and the central concentration of baryons relative to the DM distribution. Moreover, the absence of a sharp transition and the presence of a continuous trend across $V_{\rm c}$ suggest that the DM fraction is primarily regulated by stellar mass and surface density (see Section~\ref{sec:Fdm-SD-Mstar}) rather than by circular velocity alone.

%%%%%%%%%%%%%%%%%%%%%%
These findings are consistent with those observed in local disk galaxies \citep[e.g.,][]{PS1996, Courteau2015}, as well as in high-$z$ star-forming, disk-like galaxies \citep{GS23, GS21b}, where it has been demonstrated that, at fixed circular velocity, lower-mass galaxies exhibit a higher DM fraction within a representative radius compared to their high-mass counterparts. Furthermore, the positions of the Milky Way, having $M_{\rm star} = 10^{10.78}\,M_\odot$ \citep[][and ref. therein]{MilkyWay} and Andromeda, with $M_{\rm star}=10^{11.01}\,M_\odot$ \citep[][and ref. therein]{Andromeda}, lie well within the locus of our most massive ($\log(M_{\rm star}/M_\odot) = 10.5-11.5$) MAGPI galaxies. That is, the $f_{_{\rm DM}}$ values of these massive high-redshift systems are broadly consistent with those observed in local massive spiral benchmarks. 
This systematic, \textit{yet relative}, evolution in the dark matter-to-baryon ratio since $z \sim 0.85$ directly mirrors macroscopic structural observations. Specifically, it aligns with kinematic studies demonstrating that the zero-point and slope of both the stellar and baryonic Tully-Fisher relations point toward a gradual, slow evolution within the optical extent of  star-forming rotation dominated galaxies over the past $\sim 8$\,Gyr \citep[e.g.,][]{Ubler2017, AT2019, GS24, ManceraPina2026, Alexandre2026}. Additionally, it provides an empirical confirmation of cosmological abundance matching models, which show that the global stellar-to-halo mass relation ($M_{*}/M_{\rm halo}$) for star-forming disk galaxies undergoes weak evolution below $z \sim 1$ \citep[e.g.,][]{Moster2013, Behroozi2019}. This implies that within the optical extent of the disk, baryonic growth and dark matter halo assembly proceed in a highly balanced, self-similar fashion across cosmic time—a hallmark of stable disk evolution that holds true even when utilizing fully resolved dynamical treatments \citep[e.g.,][]{ManceraPina2026}.

Although the scatter in $f_{_{\rm DM}}$ at fixed $V_{\rm c}$ reflects a genuine diversity in baryon-to-DM ratios among galaxies with similar rotation velocities—potentially capturing differences in formation histories, feedback mechanisms, or baryonic concentrations—approximately 25\% of this scatter can be attributed to measurement uncertainties in $f_{_{\rm DM}}$. Beyond these errors, this dispersion is driven primarily by disk-size evolution \citep[cf.][see also \cite{GS21b, GS22}]{GS21a} and secondarily by progenitor bias. At the lower-mass end, galaxies frequently drop out of our sample due to low surface brightness limits and low signal-to-noise ratios \citep[see also][]{VanDokkum2001, LillyCarollo2016}, the so-called Tolman Dimming effect, a
relativistic phenomenon in which the observed surface brightness of a celestial object diminishes as $(1+z)^2$ \citep{Tolman1930, Tolman_diming_1996, Tolman_diming_2010}, making it more difficult to measure flux from fainter sources. 
Conversely, at the high-mass end, classic progenitor bias dominates due to galaxy quenching; as highlighted by \cite{ManceraPina2026}, the most massive, highly star-forming disks observed at $z \sim 1$ are statistically destined to quench over cosmic time, causing them to transition into passive populations that drop out of star-forming disk catalogs at lower redshifts ($z < 1$). 
Hence, this systematic shifting of population boundaries across epochs introduces an artificial dispersion when directly comparing samples from different cosmic times.
%(even across narrow redshift ranges). %The absence of a sharp transition and the presence of a continuous trend across $V_{\rm c}$ suggest that the DM fraction is primarily regulated by stellar mass and surface density (see Section~\ref{sec:Fdm-SD-Mstar}) rather than by circular velocity alone.

%###########
Furthermore, we remark that a very small subset of galaxies, in Figure~\ref{fig:Fdm-Vc-Mstar}, appear to exhibit negative DM fractions (i.e., \( f_{_{\rm DM}} < 0 \)) within \( R_{\rm e} \) (3 galaxies), particularly at the high-mass end. These unphysical values suggest that the baryonic mass, inferred from a combination of photometric SED fitting and gas scaling relation, the enclosed dynamical mass, within the characteristic radius ($R_e$ or $R_{opt}$), inferred from rotation curves cannot be less than the total observed baryonic mass. While such discrepancies could arise from systematics in the kinematic modelling, we attribute the bulk of this effect to systematics in stellar mass estimation. Specifically, recent comparisons with stellar masses derived from full-spectral fitting (see Appendix~\ref{sec:Mstar2}) indicate that braod-band SED-fitting based stellar masses tend to be systematically higher (see Figure~\ref{fig:Mstar2-fdm2}).  Applying full-spectral fitting (Poci et al. in prep.) significantly reduces or eliminates the population of ``dark matter-deficient ($f_{_{\rm DM}}<20\%$)'' galaxies, bringing their \( f_{_{\rm DM}} \) values into physically plausible ranges, see Figure~\ref{fig:Mstar2-fdm2}. This reinforces the robustness of our overall trends and highlights the importance of more precise stellar mass estimation in studies of DM content. We note that the full-spectral fitting stellar masses are not yet available for full sample used in this work; therefore, we adopt the photometric SED-based stellar masses as our baseline for all primary analyses presented in this study.

\subsection{Dark matter dynamics: scaling $f_{_{\rm DM}}$ with $\Sigma_{\rm bar}$}
\label{sec:Fdm-SD-Mstar}
\begin{figure*}[h]
	\begin{center}
\includegraphics[angle=0,height=5.0truecm,width=14.0truecm]{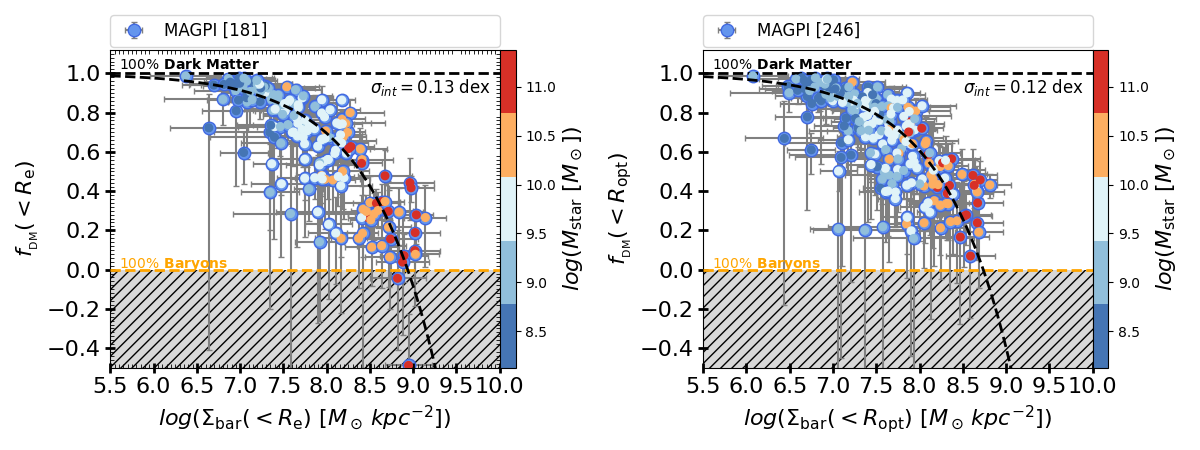} 
\caption{DM fraction (\( f_{_{\rm DM}} \)) as a function of baryon mass surface density (\( \Sigma_{\rm bar}\)). 
{\em Left and right panels:} DM fraction within the effective radius (\( R_{\rm e} \)) and optical radius (\( R_{\rm opt} \)), respectively. In both panels, individual galaxies are color-coded by stellar mass (\( M_{star} \)). The errors on the datasets are 68\% confidence intervals. 
The black dashed curve shows the best-fit relation, and intrinsic scatter (\( \sigma_{\rm int} \)) around best-fit is printed at the top right corner. The black and orange horizontal dashed lines mark the 100\% DM and 100\% baryon limits, respectively. The gray hatched region corresponds to unphysical (negative) DM fractions. Both panels highlight the inverse correlation between the DM fraction and baryonic mass surface density, with lower surface density systems being increasingly DM dominated.}
\label{fig:Fdm-Sigma-Mstar}
\end{center}
\end{figure*}

Figure~\ref{fig:Fdm-Sigma-Mstar} shows the \( f_{_{\rm DM}} \) as a function of baryon mass surface density, measured within the effective radius (left) and out to the optical radius (right). Each point represents a MAGPI galaxy, colour-coded by stellar mass; the error bars show $1\sigma$ uncertainties. The dashed black curve indicates a power-law fit:
\begin{equation}
\label{eq:sigma_bar}
f_{_{\rm DM}}(<R) = 1 - 10^{\left( -0.13 + 0.55 \left( \log \frac{\Sigma_{\rm bar}}{[M_\odot \ kpc^{-2}]} - \alpha \right) \right)},
\end{equation}
where \( \Sigma_{\rm bar}\) is baryon mass surface density defined as \( \Sigma_{\rm bar} (<R) = \frac{M_{bar}(<R)}{\pi\,R^2} \), \( \alpha = 8.7\pm 0.1 \) for \( R_{\rm e} \) and \( \alpha = 8.5\pm 0.1 \) for \( R_{\rm opt} \). 

%The low intrinsic scatter of 0.12 dex and 0.11 dex for \( R_{\rm e} \) and \( R_{\rm opt} \), respectively, indicates that \( f_{_{\rm DM}} \) is tightly connected with \( \Sigma_{\rm bar} \).  

%%%%%%%%%%%%%%%%%%%%%%
%In both panels, the numbers 274 and 397 indicate the total number of galaxies plotted that have scale radii larger than the PSF FWHM.
%%%%%%%%%%%%%%%%%%%%%%

These results reveal a clear inverse correlation between DM fraction and baryon surface density. Galaxies with higher central baryon surface densities $\Sigma_{\rm bar}\gtrsim 10^{8.0}\, M_\odot\, kpc^{-2}$ are more baryon-dominated in their inner regions. In contrast, systems with low baryonic surface densities $\Sigma_{\rm bar} < 10^{8.0} \, M_\odot\, kpc^{-2}$ are increasingly dominated by dark matter. The low intrinsic scatter of 0.13 dex and 0.12 dex for \( R_{\rm e} \) and \( R_{\rm opt} \), respectively, indicates that \( f_{_{\rm DM}} \) is tightly connected with \( \Sigma_{\rm bar} \). In comparison to high-$z$ disk-like galaxies ($z>0.8-2.2$), studied in \citetalias{GS23}, the relation presented here is relatively tight, showing factor of two lower intrinsic scatter, suggesting that the connection between baryonic surface density and DM fraction is well established at the intermediate redshifts probed by MAGPI survey at $ 0.1 \lesssim z \lesssim 0.85$. This comparatively tight correlation could also be due to the relatively high spatial resolution ($\sim 3 $ kpc) in MAGPI dataset obtained with MUSE, which wasn't the case for \citetalias{GS23} sample ($\sim 4 $ kpc). Nevertheless, this tight correlation supports a scenario in which disk galaxy mass assembly—and the resulting equivalence between baryons and dark matter—is largely in place by $ z \lesssim 0.85$, echoing the structural regularities (e.g., flat rotation curves) observed in nearby disk galaxies. Interestingly, our findings agree with those in the local Universe, where late-type spiral galaxies and low-surface-brightness galaxies exhibit a similar dependence of the DM content on baryonic surface density \citep[e.g.,][]{deBlok1997, Dutton2007, Courteau2015, Lelli2016}

%\footnote{In terms of physical units, $0.2\ arcsec \approx 0.9 \ {\rm kpc}$ at $z=0.3$ and $\sim 0.5\ arcsec $ is $4 \ {\rm kpc}$ at $z=1$, i.e., the higher-$z$ sample is about 4 times lower spatial resolution.}

The observed increase of $\Sigma_{\rm bar}$ with stellar mass and decreasing DM  fraction in both the inner ($R_{\rm e}$) and outer ($R_{\rm opt}$) regions also support the scenario described in the introduction. In this framework, massive galaxies initially undergone more efficient gas in-fall, central condensation, and star formation, most-likely tied to the low specific angular momentum, filamentary gas accretion \citep{Catelan96a,Danovich15, Breda2018, Waterval25}, potentially followed by compaction of the gas that can lead to starbursts \citep{Dekel14,Zolotov15}. This naturally leads to concentrated gas and stellar components that are characterised by high $\Sigma_{\rm bar}$ and low DM  fractions.

In contrast, and according to the ``downsizing'' picture, lower-mass galaxies begin to assemble later, at epochs when gas accretion in a $\Lambda$CDM universe is expected to carry higher specific angular momentum \citep{Catelan96a}, causing the gas to settle at larger radii. In addition, the shallower potential wells of low-mass systems make feedback more efficient at expelling baryons (see \cite{Vogelsberger20} for a review), thereby reducing $\Sigma_{\rm bar}$ and exacerbating the difference in the DM fraction within both $R_{\rm e}$ and $R_{\rm opt}$ relative to massive galaxies.
In this context, the tight correlation we observe between DM fraction and $\Sigma_{\rm bar}$ can be interpreted as a direct imprint of the baryon assembly history, encapsulating how the timing, efficiency, and spatial distribution of baryonic growth regulate the inner-to-outer mass structure of star-forming disk (or disk-like) galaxies.

%Galaxies with higher $\Sigma_{\rm bar}$ have undergone more efficient condensation of baryonic matter into their central regions, likely tied to dissipative processes and angular momentum redistribution during disc formation \citep{Fall1980, Mo1998}.

\subsection{Dark matter dynamics: scaling $f_{_{\rm DM}}$ with $z$}
\label{sec:Fdm-z-Mstar}
\begin{figure*}[h]
	\begin{center}
\includegraphics[angle=0,height=5.0truecm,width=14.0truecm]{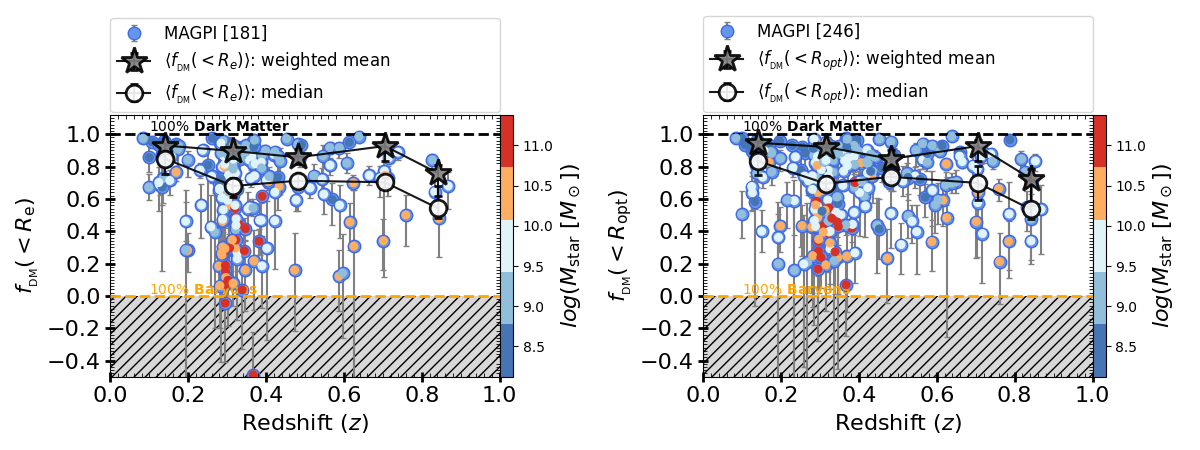} 
\caption{DM fraction (\( f_{_{\rm DM}} \)) as a function of redshift.
\textit{Left and right panels:} DM fraction within the effective radius (\( f_{_{\rm DM}}(<R_{\rm e}) \)) and optical radius (\( f_{_{\rm DM}}(<R_{\rm opt}) \)), respectively, plotted against redshift. In both panels, individual galaxies are color-coded by stellar mass (\( M_{star} \)). 
Black circles and stars represent the median and weighted mean \( \langle{f_{_{\rm DM}}} \rangle \) in redshift bins, respectively. The errors on the datasets (individual and average) are 68\% confidence intervals. Horizontal dashed lines indicate the 100\% DM (black) and 100\% baryon (orange) limits, while the gray hatched region marks unphysical negative DM fractions. These panels show majority of galaxies are DM dominated and seems to have a very mild evolution in DM fraction across the redshift range probed, with substantial scatter at all redshifts.
%In both panels, the numbers 274 and 397 indicate the total number of galaxies plotted that have scale radii larger than the PSF FWHM.
}
\label{fig:Fdm-Redshift-Mstar}
\end{center} 
\end{figure*}

Figure~\ref{fig:Fdm-Redshift-Mstar} shows the DM fraction as a function of redshift. The left panel presents \( f_{_{\rm DM}}(<R_{\rm e}) \), and the right panel shows \( f_{_{\rm DM}}(<R_{\rm opt}) \). To illustrate the relationship between baryons at different redshifts, individual galaxies are colour-coded by stellar mass. To characterise the global trend, we compute the weighted mean (grey stars) and median (white circles) of \( f_{_{\rm DM}} \) in redshift bins. In both cases, uncertainties on the binned data points are estimated via bootstrap resampling and represent the 68\% confidence interval. For details of the binning procedure, we refer the reader to \citetalias{GS23} (Sec.~4.2).
%%%%%%%%%%%%%%%%%%%%%%
%%%%%%%%%%%%%%%%%%%

%Statistically, over the MAGPI sample redshift range $0.1 \leq z \leq 0.85$, the median DM fraction within the $R_e$ is approximately 0.65, and the weighted mean is slightly higher ($\sim 0.8$). 
Statistically, over the MAGPI sample redshift range $0.1 \leq z \leq 0.85$, the median dark matter fraction within $R_{\rm e}$ is $0.70$, while the error-weighted mean is $0.90$. This shift reflects a distinct skewness in the distribution toward higher dark matter fractions, alongside smaller measurement uncertainties for dark-matter-dominated systems.
At the optical radius ($R_{\rm opt}$), the typical (average of median and weighted mean) DM fraction is about 0.85. Given the uncertainties on the measurements, within both the $R_e$ and $R_{\rm opt}$, the data show no significant trend with redshift. This suggests that for rotation-supported systems that lie on the star-forming main sequence, the balance between baryonic and DM components within these characteristic radii was already largely in place by $z \sim 0.1 -  0.8$. The lack of significant evolution is consistent with the findings of \citetalias{GS23}, who showed that high-redshift ($z \gtrsim 1$)  star-forming galaxies typically exhibit elevated velocity dispersions and lower rotation-to-dispersion ratios, but once corrected for pressure support, their DM fractions within the inner regions are comparable to those of local disks \citep{PS1996,Courteau2015}. The MAGPI survey data now empirically support this result for intermediate redshifts. 

The vertical scatter in \( f_{_{\rm DM}} (<R_e) \) within each redshift bin is driven by stellar mass and progenitor bias. Consistent with above discussions, we observe that lower-mass disks (\( M_{\rm star} < 10^{9.5}~M_\odot \)) are generally more dark-matter dominated within their inner regions, whereas the  massive systems (\( M_{\rm star} \geq 10^{10.0}~M_\odot \)) tend to be baryon-dominated at their centers. If a complete stellar mass range (\( \log(M_{\rm star} [M_\odot]) = 8.0{-}11.5 \)) were uniformly sampled across all redshift intervals, it is likely that this mass-dependent trend would be apparent at all redshifts. However, we do not explicitly observe this because low-mass galaxies are intrinsically difficult to detect at high redshift; those that are observed typically have lower signal-to-noise ratios, resulting in larger uncertainties in both dynamical and baryonic mass estimates, and consequently, in the inferred dark matter fractions. In contrast, the scatter due to mass is less pronounced in the outer regions, $f_{_{\rm DM}}(<R_{\rm opt})$, we explore this more in Section~\ref{sec:FdmEvo} where we combine MAGPI sample with \citetalias{GS23} high-$z$ sample.

The flat $f_{_{\rm DM}}(z)$ relation suggests that, although galactic disks have grown in stellar mass and size by a factor of $\sim 2$ since $z \sim 1$, the accompanying accretion of DM and the inside‐out buildup of baryons have proceeded in lockstep \citep[see also][and references therein]{VanDokkum2013}. Moreover, the substantial scatter in \( f_{_{\rm DM}} \) at fixed redshift and radius most-likely reflects the diversity of evolutionary pathways, including differences in gas accretion, star formation efficiency, and feedback history across the galaxy population.

%%%%%%%%%%%%%%

\subsection{Statistical relevance of various correlations}
\label{sec:fdm-stats}
Table~\ref{tab:correlation_stats} presents the Pearson correlation coefficients ($r$) and associated $p$‑values for the various scaling relations analysed in this study. Statistically, the coefficient $r$ quantifies the strength and direction of a linear relationship: values near +1 indicate a strong positive correlation, those near –1 indicate a strong negative correlation, and values near zero indicate no significant correlation. The corresponding $p$‑value assesses statistical significance; we remark that $p<0.05$ is used as the threshold for a statistically robust correlation. These correlations are computed over the UIDs (including weighted values of repeated objects) for DM fraction within both $R_{\rm e}$ and $R_{\rm opt}$. %We also investigate the relationship between redshift and $f_{_{\rm DM}}$ within narrow stellar mass bins, restricting to $z=0.2$–0.5, where the full stellar mass range ($8.0\leq \log(M_{\rm star}\,[M_\odot]) \leq 11.5)$ is represented (see Figure~\ref{fig:Fdm-Redshift-Mstar}).

\begin{table}[ht]
\centering
\begin{tabular}{l|c|c|c|c}
\hline
\textbf{Relation Name} & bins & \textbf{Npoints} & \textbf{Corr. Coeff.} & \textbf{p-value} \\
\hline
$V_c$ vs $f_{_{\rm DM}}(<R_e)$  & -  & 180 &  $-0.4036$ & $1.76e-08$  \\
$V_c$ vs $f_{_{\rm DM}}(<R_{\rm opt})$ & - & 245 &  $-0.1792$ & $0.0048$\\
\hline 
$\Sigma_{\rm bar}$ vs $f_{_{\rm DM}}(<R_e)$  & -  & 180 &  $-0.8070$ & $<1e-10$\\
$\Sigma_{\rm bar}$ vs $f_{_{\rm DM}}(<R_{\rm opt})$ & - & 245 &  $-0.7352$ & $<1e-10$ \\
\hline
$M_{\rm star}$ vs $f_{_{\rm DM}}(<R_e)$  & -  & 180 &  $-0.7294$ & $<1e-10$\\
$M_{\rm star}$ vs $f_{_{\rm DM}}(<R_{\rm opt})$ & - & 245 &  $-0.5819$ & $<1e-10$\\
\hline
$z$ vs $f_{_{\rm DM}}(<R_e)$  & -  & 180 &  $0.0017$ & 0.9814 \\
$z$ vs $f_{_{\rm DM}}(<R_{\rm opt})$ & - & 245 &  $0.0029$ & 0.9636\\
\hline
\end{tabular}
\caption{Correlation statistics for various scaling relations. We present Pearson correlation coefficients ($r$) and associated $p$-values for DM fraction ($f_{_{\rm DM}}$) as a function of different galaxy properties within the effective radius ($R_{\rm e}$) and optical radius ($R_{\rm opt}$). Correlations with $p{\rm -value} < 0.05$ are considered statistically significant. Strong negative correlations are observed between $f_{_{\rm DM}}$ and both $\Sigma_{\rm bar}$ and $M_{star}$, indicating that higher baryonic surface density and stellar mass are associated with lower DM fractions, particularly within $R_{\rm e}$. Correlations with $V_c$ and redshift are also evaluated.
}
\label{tab:correlation_stats}
\end{table}

We find strong negative correlations between $f_{_{\rm DM}}$ and both baryonic surface density $\Sigma_{\rm bar}$ and stellar mass $M_{\rm star}$, within $R_{\rm e}$ and $R_{\rm opt}$. These relations are highly significant (all $p$-values $< 10^{-10}$), with $r$ reaching below $-0.5$, especially in the inner regions, suggesting that galaxies with higher stellar mass and central baryonic concentration tend to have lower DM fractions in their inner regions. These findings align with the trends discussed in Sections~\ref{sec:Fdm-Vc-Mstar} and \ref{sec:Fdm-SD-Mstar}. 
Circular velocity, $V_c$, exhibits a moderate negative correlation with $f_{_{\rm DM}}(<R_{\rm e})$ ($r=-0.4$, $p=1.8\times10^{-8}$), suggesting that faster-rotating galaxies are generally more baryon-dominated in their centers compared to outskirts. In contrast, no significant correlation is found between $V_c$ and $f_{_{\rm DM}}(<R_{\rm opt})$, implying that DM remains quite constant in the outskirts regardless of rotation speed (see Section~\ref{sec:Fdm-Vc-Mstar}). We also observe a statistically insignificant negative correlation between redshift and $f_{_{\rm DM}}$ ($r\approx 0.002$ within $R_{\rm e}$ and $R_{\rm opt}$), which we have further discussed in Section~\ref{sec:FdmEvo}.

\section{Dark matter evolution with cosmic-time}
\label{sec:FdmEvo}
\begin{figure*}[h]
	\begin{center}
\includegraphics[angle=0,height=4.5truecm,width=14.0truecm]{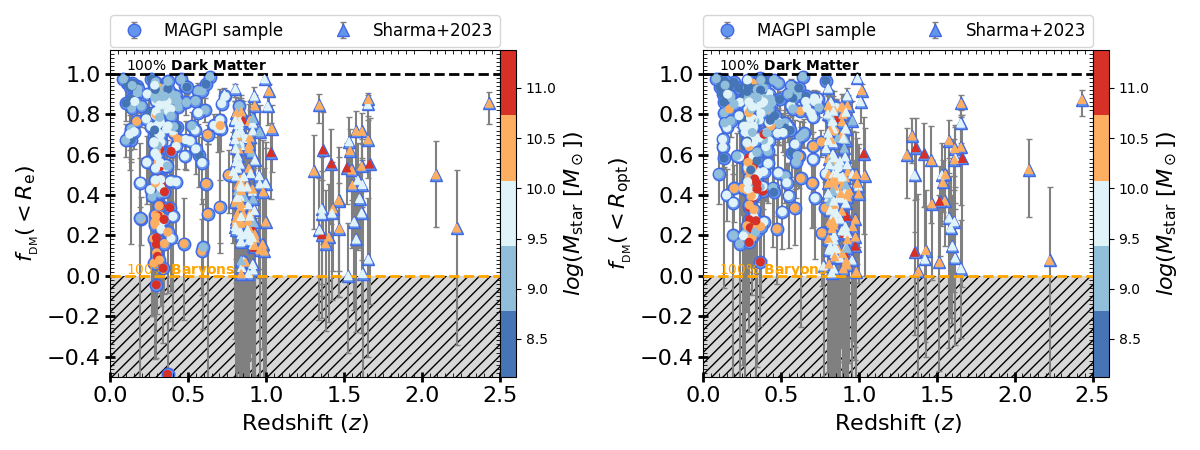} 
\includegraphics[angle=0,height=6.5truecm,width=9.0truecm]{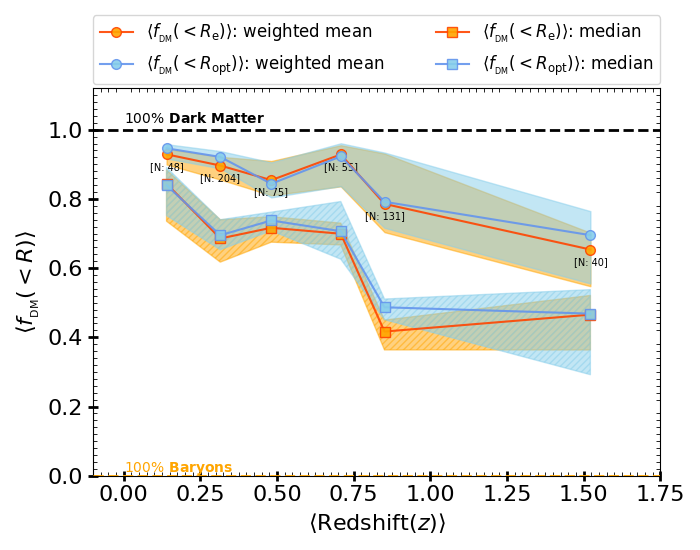}
\caption{DM fraction (\( f_{_{\rm DM}} \)) as a function of redshift for the MAGPI sample and high-redshift galaxies from \citetalias{GS23}. 
\textit{top left panel:} \( f_{_{\rm DM}}(<R_{\rm e}) \) versus redshift. 
\textit{top right panel:} \( f_{_{\rm DM}}(<R_{\rm opt}) \) versus redshift. MAGPI galaxies are shown as circles while \citetalias{GS23} galaxies are shown as triangles, both datasets are color-coded by stellar mass. {\em Bottom panel:}  Median values are indicated by square-connected solid lines with hatched shaded regions, and weighted mean values by circle-connected solid lines with plain shaded regions. The shaded regions are showing $1\sigma$ uncertainty, where 
Orange represents \( f_{_{\rm DM}}(<R_{\rm e}) \), and blue represents \( f_{_{\rm DM}}(<R_{\rm opt}) \).}
\label{fig:Fdm-Redshift-MP-GS23}
\end{center}
\end{figure*}

To obtain a comprehensive view of the DM content across cosmic time ($z=0.1 - 1.5$) for the \textit{star-forming disk galaxy population} ($log(M_{star}\ [M_\odot]) = 8.0-11.5$), we combined the MAGPI sample with high-redshift data from \citetalias{GS23}. The latter dataset was processed and analysed using the same modelling tools, techniques, and methodology applied in this work. In Figure~\ref{fig:Fdm-Redshift-MP-GS23}, we present combined dataset showing the trend of the DM fraction with redshift. The top-left and top-right panels show \( f_{_{\rm DM}}(<R_{\rm e}, z) \) and \( f_{_{\rm DM}}(<R_{\rm opt}, z) \), respectively, with individual galaxies color-coded by their stellar mass. The bottom panel summarizes the median and weighted mean \( f_{_{\rm DM}} \) trends for both radii, with shaded regions representing the 68\% confidence intervals. We notice that across the MAGPI redshift window the typical dark-matter fraction remains nearly constant, with $\langle f_{_{\rm DM}}(<R_{\rm e})\rangle\simeq0.75\pm 0.1$ and $\langle f_{_{\rm DM}}(<R_{\rm opt})\rangle\simeq0.82 \pm 0.1$ (also see Figure~\ref{fig:Fdm-Redshift-Mstar}). At $z\gtrsim1$ the \citetalias{GS23} galaxies exhibit lower values, producing monotonic decline in \( f_{_{\rm DM}}(<R) \) with redshift. This declining trend is mostlikely due to a observational bias, the so-called Tolman Dimming effect, a
relativistic phenomenon in which the observed surface brightness of a celestial object diminishes as $(1+z)^2$ \citep{Tolman1930, Tolman_diming_1996, Tolman_diming_2010}, making it more difficult to measure flux from fainter sources. The smaller sizes of low-mass galaxies is an additional challenge for resolving full rotation curves. 
%Consequently, low mass or low brightness galaxies could go missing in high-redshift observations.

%%%%%%%%%%%%%%%%%%%%%%

%%%%%%%%%%%%%%%%%%%%%%
%To assess the impact of this observational bias on our measurements, we divide the combined galaxy sample into four stellar mass bins and examine their DM fractions within $R_{\rm e}$ and $R_{\rm opt}$, as shown in Figures~\ref{fig:Fdm-Redshift-Mstar_bins_Re} and \ref{fig:Fdm-Redshift-Mstar_bins_Ropt}, respectively.

%%%%%%%%%%%%%%%%%%%%%%
\begin{figure*}[h]
	\begin{center}
\includegraphics[angle=0,height=7.5truecm,width=13.0truecm]{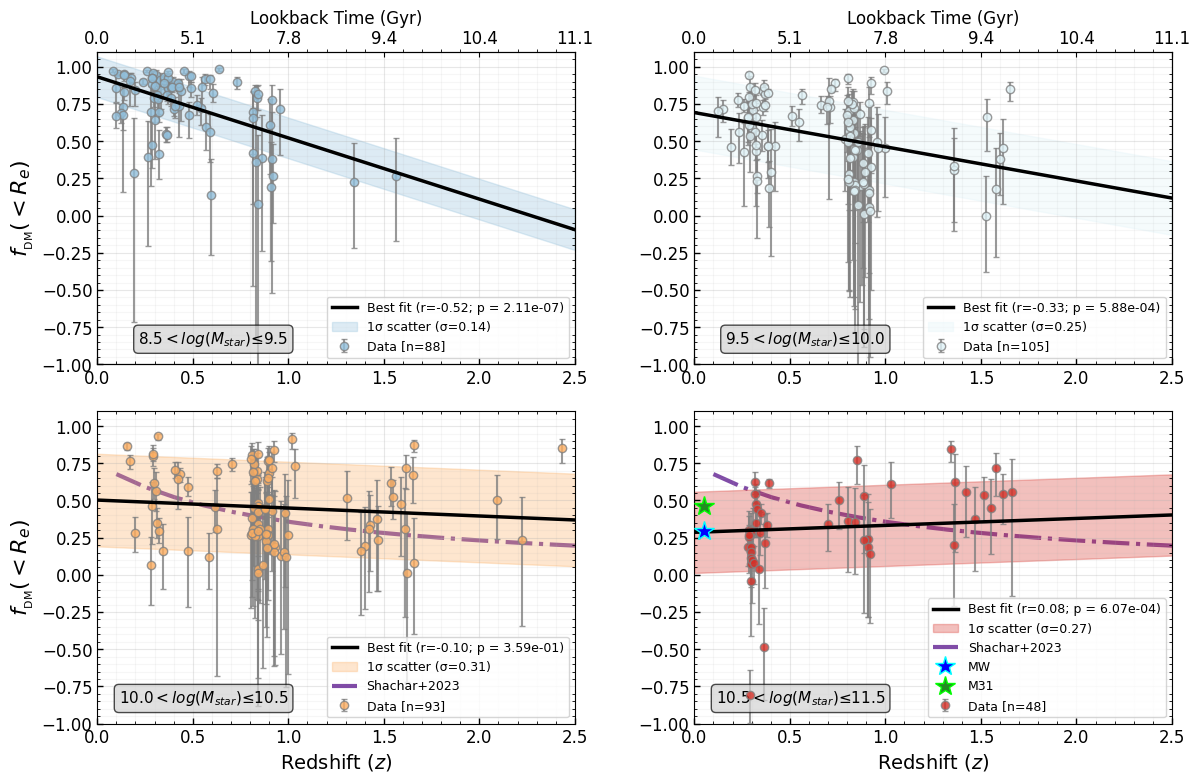} 
    \caption{Redshift evolution of DM fraction within the effective radius, $f_{_{\rm DM}}(<R_{\rm e})$, in four stellar mass bins using the combined MAGPI+GS23 sample. Each panel corresponds to a different stellar mass bin: 
        \textit{Top-left}: $8.5 < \log(M_{star}\ [M_\odot]) \leq 9.5$, 
        \textit{Top-right}: $9.5 < \log(M_{star}\ [M_\odot]) \leq 10.0$, 
        \textit{Bottom-left}: $10.0 < \log(M_{star}\ [M_\odot]) \leq 10.5$, and 
        \textit{Bottom-right}: $10.5 < \log(M_{star}\ [M_\odot]) \leq 11.5$. 
        The $y$-axis shows $f_{_{\rm DM}}(<R_{\rm e})$, while the $x$-axis denotes redshift (bottom axis) and lookback time (top axis). Each data point represents a galaxy, with error bars indicating observational uncertainties. The black lines represent best-fit linear regressions and shaded region shows the 1$\sigma$ intrinsic around best-fit. Pearson correlation coefficients ($r$) and $p$-values are printed in the legend alongside best-fit. The purple dash-dotted line indicates the results from \cite{Genzel2022} for a comparable stellar mass range. For reference, the DM fractions within $R_{\rm e}$ for the Milky Way (MW) and Andromeda (M31) are shown as blue and green stars, respectively, representing massive local star-forming disk galaxies.
    }
\label{fig:Fdm-Redshift-Mstar_bins_Re}
\end{center}
\end{figure*}

\begin{figure*}[h]
	\begin{center}
\includegraphics[angle=0,height=7.5truecm,width=13.0truecm]{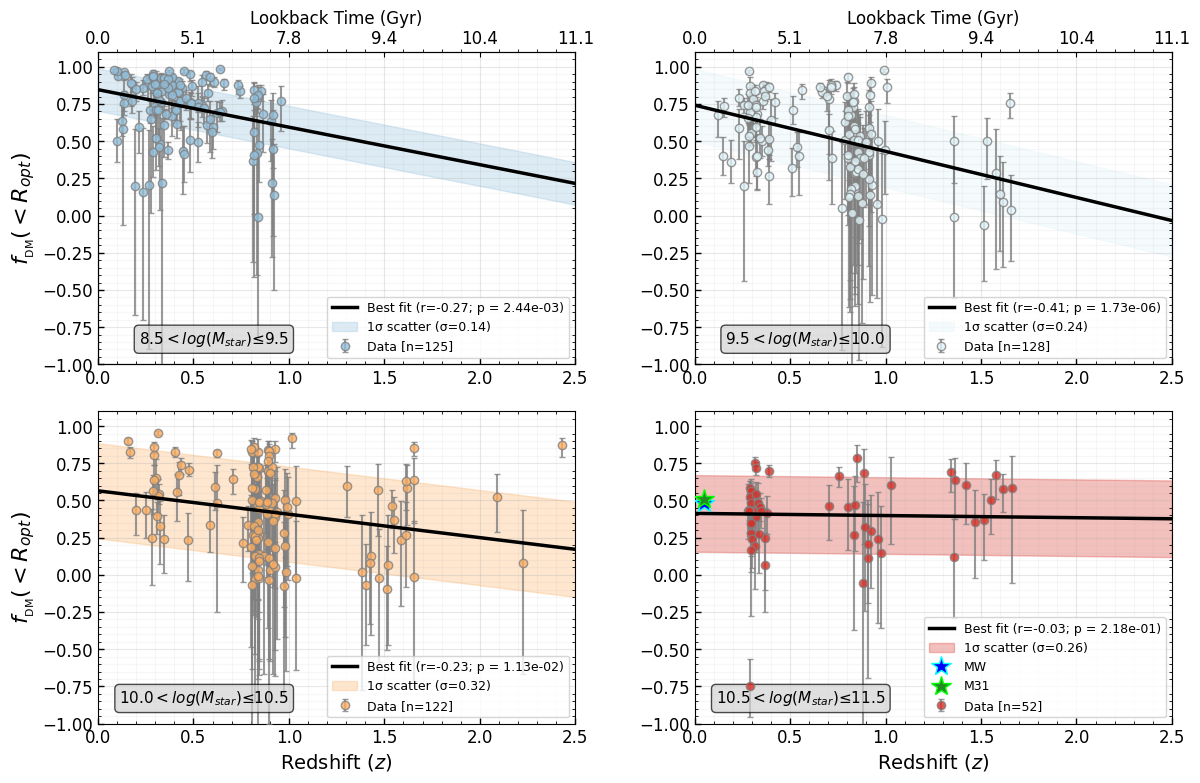} 
    \caption{Redshift evolution of DM fraction within the optical radius, $f_{_{\rm DM}}(<R_{\rm opt})$, in four stellar mass bins using the combined MAGPI+GS23 sample. Each panel corresponds to a different stellar mass bin: 
        \textit{Top-left}: $8.5 < \log(M_{star}\ [M_\odot]) \leq 9.5$, 
        \textit{Top-right}: $9.5 < \log(M_{star}\ [M_\odot]) \leq 10.0$, 
        \textit{Bottom-left}: $10.0 < \log(M_{star}\ [M_\odot] \leq 10.5$, and 
        \textit{Bottom-right}: $10.5 < \log(M_{star}\ [M_\odot]) \leq 11.5$. 
        The $y$-axis shows $f_{_{\rm DM}}(<R_{\rm opt})$, while the $x$-axis denotes redshift (bottom axis) and lookback time (top axis). Each data point represents a galaxy, with error bars indicating observational uncertainties. The black lines represent best-fit linear regressions and shaded region shows the 1$\sigma$ intrinsic around best-fit. Pearson correlation coefficients ($r$) and $p$-values are printed in the legend alongside best-fit. For reference, the DM fractions within $R_{\rm opt}$ for the Milky Way (MW) and Andromeda (M31) are shown as blue and green stars, respectively, representing massive local star-forming disk galaxies.
    }
\label{fig:Fdm-Redshift-Mstar_bins_Ropt}
\end{center}
\end{figure*}
%%%%%%%%%%%%%%%%%%%%%%

%Due to differences in targeting strategy between the KMOS and MUSE surveys, the stellar mass ranges probed are different. Therefore, to

To assess the impact of this observational bias on our measurements, we divide the combined galaxy sample into four stellar mass bins and examine their DM fractions within $R_{\rm e}$ and $R_{\rm opt}$, as shown in Figures~\ref{fig:Fdm-Redshift-Mstar_bins_Re} and \ref{fig:Fdm-Redshift-Mstar_bins_Ropt}, respectively. We find that low-mass galaxies ($\log(M_{star}\ [M_\odot]) = 8.5\text{--}9.5$) are absent at $z > 0.9$, and only a small number ($\sim$10) of intermediate-mass galaxies ($9.5 < \log(M_{star}\ [M_\odot]) \leq 10.0$) provide robust measurements at $z \sim 1.5$. As a result, when averaging the DM fraction above $z > 0.9$ (see Figure~\ref{fig:Fdm-Redshift-MP-GS23} bottom panel), the results are biased toward massive galaxies ($\log(M_{star} [M_\odot]) > 10$) which tend to show relatively low dark matter fractions.

Within the stellar mass range \(8.5 < \log(M_{star}/M_\odot) \leq 10.5\), galaxies tend to show a systematic decrease in dark matter fraction with increasing redshift, both within \(R_{\rm e}\) and \(R_{\rm opt}\). For the lowest–mass bin \(8.5 < \log(M_{star}/M_\odot) \leq 9.5\), we find a relatively strong anti–correlation between \(f_{_{\rm DM}}(<R_{\rm e})\) and redshift (top–left panel; \(r = -0.52\), \(p = 2.1\times10^{-7}\)) and a moderate anti–correlation for \(f_{_{\rm DM}}(<R_{\rm opt})\) (top–left panel; \(r = -0.27\), \(p = 2.44\times10^{-3}\)). In the second bin, \(9.5 < \log(M_{star}/M_\odot) \leq 10.0\), the trends remain significant but shallower, with \(r = -0.33\) (\(p = 5.9\times10^{-4}\)) for \(f_{_{\rm DM}}(<R_{\rm e})\) and \(r = -0.41\) (\(p = 1.7\times10^{-6}\)) for \(f_{_{\rm DM}}(<R_{\rm opt})\). For intermediate–mass systems, \(10.0 < \log(M_{star}/M_\odot) \leq 10.5\), the correlations weaken: \(r = -0.10\) (\(p = 0.36\)) within \(R_{\rm e}\) and \(r = -0.23\) (\(p = 1.1\times10^{-2}\)) within \(R_{\rm opt}\). In all three bins, the data at the high-redshift end ($z > 1$) display larger individual uncertainties. If we interpret these continuous trends as a signature of cosmic evolution, they reflect a complex interplay between global halo assembly and internal baryonic redistribution. On one hand, this behavior provides direct empirical evidence of hierarchical growth, where galaxies continuously accrete dark matter onto their halos over cosmic time. On the other hand, because $f_{_{\rm DM}}$ is measured within fixed physical apertures ($R_{\rm e}$ or $R_{\rm opt}$), this signature is also impacted by the rapid structural evolution processes of low-to-intermediate-mass systems.

In contrast, the most massive galaxies, with \(10.5 < \log(M_{\star}/M_\odot) \leq 11.5\), show no statistically significant evolution in dark matter fraction within \(R_{\rm opt}\) (bottom–right panel; \(r = 0.08\), \(p = 0.2\)), and only a mild positive trend within \(R_{\rm e}\) (bottom–right panel; \(r = -0.03\), \(p = 6.07\times10^{-4}\)), which remains small compared to the intrinsic scatter. Their mean dark matter fractions, \(\langle f_{_{\rm DM}}(<R_{\rm e}) \rangle \approx 0.40\) and \(\langle f_{_{\rm DM}}(<R_{\rm opt}) \rangle \approx 0.47\), are approximately constant over the full redshift range and comparable to those measured in massive local disks such as the Milky Way and M31. The lack of strong evolution at high masses suggests that (1) progenitors of present‑day massive systems may have evolved in a manner analogous to their high‑redshift counterparts, and (2) galaxies might evolve in such a way that increases in baryonic mass and growth of dark matter halo proceed in a co‑regulated fashion so that the inner and outer DM fraction remains approximately constant.

As is visually apparent in Figures~\ref{fig:Fdm-Redshift-Mstar_bins_Re} and \ref{fig:Fdm-Redshift-Mstar_bins_Ropt}, the measurement uncertainties on DM fraction systematically increase for the highest-$z$ galaxies in \citetalias{GS23} sample. This reflects a rigorous Monte Carlo error propagation that accounts for three fundamental observational realities at high redshift. First, cosmological surface brightness dimming reduces the signal-to-noise ratio in the extended disk, widening the posterior parameter space during the kinematic modeling. Second, the larger physical scale of the PSF at high $z$ leaves fewer independent spatial resolution elements, exacerbating morphological degeneracies during the baryonic mass decomposition and kinematic modelling. Third, the high-$z$ baryonic mass budgets are dominated by larger systematic uncertainties; high-$z$ stellar masses ($M_\star$) are often drawn from diverse surveys utilizing varying SED-fitting techniques, whereas the lower-redshift MAGPI stellar masses are derived uniformly using one single SED-fitting technique. Consequently, the enlarged error bars properly reflect the degraded data quality and systematic offsets at high redshift, ensuring that our finding of no statistically significant redshift evolution remains a robust, accurately weighted conclusion.

%%%%%%%%%%%%%%%%%%%%%%%%%%%%%%%%%%%%%%%%%%%%%%%%%%%%%%%%%%

%\section{Conclusion}\label{sec:conclusion}
To conclude, due to the targeting strategy of MAGPI, similar to \cite{Ciocan2025}, our sample is dominated by low- to intermediate-mass galaxies ($\log(M_{\rm star}/M_\odot) < 10$; Figure~\ref{fig:MAGPI-hist}). At these mass scales, we find high dark matter fractions within $R_{\rm e}$ ($\gtrsim 50\%$), consistent with MUSE-DARK survey \cite{Ciocan2025}, the KURVS survey \citep{KURVS}, and KMOS$^{\rm 3D}$ \cite{Genzel2022}. When all galaxies in our combined sample (MAGPI+\citetalias{GS23}) are considered simultaneously, we do not detect a statistically significant trend with redshift (Figure~\ref{fig:Fdm-Redshift-Mstar} \& \ref{fig:Fdm-Redshift-MP-GS23}), implying little global evolution across the general population ($8.0 \leq log(M_{\rm star}) \leq 11.5$). However, when the sample is dissected by stellar mass, a distinct mass-dependence emerges: low-mass galaxies exhibit the strongest variation in their, both inner and outskirts, DM fractions over time, whereas high-mass systems show virtually no evolution. In the intermediate-to-high mass bin ($10 < \log(M_{\rm star}/M_\odot) \leq 10.5$), our results are structurally consistent with the weak evolution measured by \cite{Genzel2022}, though the statistical significance of this trend within our standalone dataset is not robust. This stark contrast across the mass baseline suggests that the physical mechanisms regulating the central mass distributions of star-forming disks operate with differing efficiencies depending on the total mass of the system.

This mass-dependent behavior can also be understood through the lens of co-regulated galaxy assembly, where internal baryonic processes scale in tandem with dark matter halo growth. From various observations and simulations, it is well established that star-forming disk galaxies undergo inside-out growth, expanding their stellar effective radii toward lower redshifts due to the late-time accretion of high specific angular-momentum gas \citep{Dutton2011, Stewart2013, vanderWel2014, Nelson2016}. Concurrently, stellar feedback acting within shallower gravitational potential wells can efficiently redistribute baryonic mass outward from the central core \citep{Brook2011}. Because the underlying dark matter halo profile is spatially extended, the continuous global accretion of dark matter in parallel with inside-out baryonic expansion naturally causes fixed physical apertures ($R_{\rm e}$ or $R_{\rm opt}$) to enclose an increasingly larger dark matter fraction at lower redshifts. This combined mechanism explains why the amplitude of this apparent temporal shift is significantly enhanced in lower-mass systems ($f_{_{\rm DM}} \approx 0.85 \rightarrow 0.5$ across our mass baseline) compared to their more massive, structurally mature counterparts; i.e., low-mass galaxies exhibit a higher degree of evolution (also cf. \citetalias{GS23}).

A flat \textit{global} $f_{_{\rm DM}}(z)$ relation, combined with the declining $f_{_{\rm DM}}(M_{\rm star})$ and rising $f_{_{\rm DM}}(R)$ trends, ultimately supports a scenario where the growth of stellar disks proceeds in parallel lockstep with global dark matter halo assembly \citep[as also suggested by][]{VanDokkum2013}.  High-resolution hydrodynamical cosmological simulations incorporating strong, mass-dependent feedback and realistic halo contraction strongly mirror these observational trends. For example, IllustrisTNG results \citep{Lovell2018} demonstrate that galaxies maintain a balanced co-evolution between baryons and DM such that their rotation curves remain flat, resulting in a roughly fixed inner dark matter fraction over cosmic time. Likewise, \cite{Wechsler2018} and EAGLE simulation frameworks \citep{Wright2020} show that incorporating stellar/AGN feedback alongside gradual halo responses yields stable, nearly constant baryon-to-DM fractions across scales. While these simulations focus heavily on massive systems—supporting the relative stability we observe at our high-mass end—our empirical findings indicate that at lower mass scales, the observational signatures of this co-evolution are more dynamically dynamic due to the changing structural sizes of the disks. In a follow-up work, we will compare our results quantitatively with mock observations derived from these simulations.

\section{Summary}
\label{sec:summary}
To investigate the co-evolution of baryonic and dark matter (DM), we have presented a population-level analysis of the dark matter fraction ($f_{_{\text{DM}}}$) and its scaling relations using a sample of 266 star-forming, disk-like galaxies at intermediate redshifts ($0.1 \le z \le 0.85$) from the MAGPI survey. This sample spans a broad stellar mass range of \(\log(M_{star}\ [M_{\odot}]) = 8.0{-}11.5\), allowing us to characterize the overall statistical behavior of the star-forming main sequence population across an epoch corresponding to cosmic ages between $\sim 6.7 -12.5$ Gyr. By performing a robust three-dimensional forward-modeling kinematic analysis with 3DBarolo and applying gas pressure gradient corrections, we constructed model-independent circular velocity profiles. Decomposing these profiles into stellar disks, spheroidal bulges, and multi-phase gas components enabled the estimation of enclosed dark matter fractions within both the stellar effective radius ($R_e$) and the optical radius ($R_{\text{opt}}$) without relying on parametric halo models. The results are presented and discussed in Section~\ref{sec:result}-\ref{sec:FdmEvo}. We also quantitatively assess the statistical relevance of all scaling relation shown in the Table~\ref{tab:correlation_stats}. 
\\
\\
The key conclusions of this work are as follows:

\begin{itemize}
\item \textbf{Mass-Dependent Structural Dichotomy:} Low-mass galaxies ($M_{\text{star}} < 10^{9.5} M_{\odot}$) are dark-matter-dominated across all radii, exhibiting a high average dark matter fraction of $\langle f_{_{\text{DM}}}\rangle \approx 0.85$ within both the inner regions ($R_e$) and the outskirts ($R_{\text{out}}$). This suggests that energetic stellar feedback and environmental effects (e.g., ram pressure stripping and tidal interactions) efficiently remove gas from shallower potential wells, rendering star formation highly inefficient, maintaining low baryonic surface densities, and leaving the gravitational potential dark-matter-dominated.  
\item \textbf{Radial Gradients in Massive Disks:} In high-mass systems ($M_{\text{star}} \ge 10^{10} M_{\odot}$), the dark matter fraction is low within the central regions ($\langle f_{_{\text{DM}}}\rangle \approx 0.47$ within $R_e$) but rises significantly toward the outskirts. At a fixed circular velocity, low-mass galaxies exhibit a central dark matter fraction that is approximately 55\% higher than their high-mass counterparts within $R_e$, whereas the outskirts of both mass regimes show relatively differences ($\sim$ 30\%). This distinct structural behavior is highly consistent with an inside-out disk growth scenario. These findings are also consistent with the predictions of local spirals and previous high-$z$ disk-like galaxies.
\item \textbf{Tight Coupling with Baryon Surface Density:} We report a strong, inverse correlation between $f_{_{\text{DM}}}$ and the baryon mass surface density ($\Sigma_{\text{bar}}$) across both the inner regions and outskirts. This relation is remarkably tight, with a low intrinsic scatter of 0.12 to 0.13 dex, mirroring the correlations established in the local Universe. This tight coupling, yet again, supports an inside-out baryon assembly scenario, where dense, low specific angular momentum gas pools centrally to form mature cores at early epochs, while higher specific angular momentum gas accretes later to build the outer disk.
\item  \textbf{Mass-Dependent Cosmic Evolution:} Within the MAGPI intermediate-redshift window, we observe no statistically significant global evolution in the dark matter fraction $f_{_{\text{DM}}}(z)$. However, when combining MAGPI with the high-redshift ($0.6 \le z \le 1.5$) sample from \citetalias{GS23}, the evolutionary trajectory over the past 10 billion years reveals a distinct mass-dependence. Low-mass galaxies ($M_{\text{star}} \le 10^{9.5} M_{\odot}$) exhibit the most dynamic variation in their dark matter fractions over cosmic time. Conversely, at high mass ($M_{star}> 10^{10} \ M_\odot$) our results are consistent with \citep{Genzel2022} but are not statistically significant. At $z > 0.75$ the sample is dominated by high mass galaxies ($M_{star}> 10^{9.5} \ M_\odot$), while the MAGPI sample is dominated by low mass galaxies ($M_{star}\leq 10^{9.5} \ M_\odot$) making the need for samples with large dynamic ranges increasingly important.
\item \textbf{Observational Biases at High Redshift:} We quantitatively demonstrate that the previously reported global decline in $f_{_{\text{DM}}}(z)$ at high redshift is heavily modulated by observational selection effects. Due to Tolman surface brightness dimming, high-redshift samples ($z > 0.75$) are severely biased toward massive, baryon-dominated galaxies, while low-mass, dark-matter-dominated systems drop out of high-redshift sample or fall in the set of low SNR galaxies.
\item \textbf{Consistency with Hydrodynamical Simulations:} The flat global $f_{_{\text{DM}}}(z)$ relation, combined with the declining $f_{_{\text{DM}}}(M_{\text{star}})$ and rising $f_{_{\text{DM}}}(R)$ trends, together with its strong dependence on $\Sigma_{\rm bar}$ and $M_{star}$, supports a scenario where the growth of stellar disks proceeds in lockstep with dark matter halo assembly. This behavior is often reproduced in state-of-the-art cosmological hydrodynamical simulations. For example, IllustrisTNG shows a balanced co-evolution where rotation curves remain flat over cosmic time, while the EAGLE simulation framework demonstrates that feedback-regulated growth and gradual halo response preserve nearly stable baryon-to-dark-matter ratios across both $R_e$ and $R_{\text{opt}}$. 
\end{itemize}
Together, these results provide strong empirical support for a scenario in which disk galaxies assemble their baryons and DM in a coordinated fashion, with the dominant mass component depending on galaxy mass, scale, and baryonic concentration— but not significantly on redshift within $z\approx 0.1 - 1.5$. Next generation high-resolution IFU surveys at $z>1$ targeting lower-mass systems will be critical for disentangling observational biases and further constraining the co-evolution of baryons and DM in galaxies.

\newpage
\acknowledgments
GS acknowledge funding from the European Union’s Horizon 2020 research and innovation programme under the Marie Sklodowska-Curie Grant agreement ID No.: 101147719. Although project is funded by the European Union, views and opinions expressed in this work are however those of the author(s) only and do not necessarily reflect those of the European Union. Neither the European Union nor the granting authority can be held responsible for them. G.S. acknowledges support from the SARAO Postdoctoral Fellowship (UID 97882), under which the MAGPI project, Dark Matter Halos of Intermediate-Redshift Galaxies, was initiated. GS acknowledge the use of the ilifu cloud computing facility – www.ilifu.ac.za, a partnership between the University of Cape Town, the University of the Western Cape, Stellenbosch University, Sol Plaatje University and the Cape Peninsula University of Technology. The ilifu facility is supported by contributions from the Inter-University Institute for Data Intensive Astronomy (IDIA – a partnership between the University of Cape Town, the University of Pretoria and the University of the Western Cape), the Computational Biology division at UCT and the Data Intensive Research Initiative of South Africa (DIRISA). GS also acknowledge the funding from IRMIA++ fellowship where this project has been continued. GS thanks Benoit Famaey, Mihael Petac, and Jonathan Freundlich for various discussions. A part of this research was supported by funding from the University of Strasbourg Institute for Advanced Study (USIAS) within the French national programme “Investment for the Future” (Excellence Initiative) IdEx-Unistra, under the leadership of Florent Renaud. 
CL, TM and CF are the recipient of the Australian Research Council Discovery Project DP210101945. Based on observations collected at the European Organisation for Astronomical Research in the Southern Hemisphere under  ESO program 1104.B-0536. We wish to thank the ESO staff, and in particular the staff at Paranal Observatory, for carrying out the MAGPI observations. MAGPI targets were selected from GAMA. GAMA is a joint European-Australasian project based around a spectroscopic campaign using the Anglo-Australian Telescope. GAMA is funded by the STFC (UK), the ARC (Australia), the AAO, and the participating institutions. GAMA photometry is based on observations made with ESO Telescopes at the La Silla Paranal Observatory under programme ID 179.A-2004, ID 177.A-3016. 
CF is the recipient of an Australian Research Council Future Fellowship (project number FT210100168) funded by the Australian Government.  
SMS acknowledges funding from the Australian Research Council (DE220100003). 
Parts of this research were conducted by the Australian Research Council Centre of Excellence for All Sky Astrophysics in 3 Dimensions (ASTRO 3D), through project number CE170100013.
I.B. has received funding from the European Union’s Horizon 2020 research and innovation programme under the Marie Sklodowska-Curie Grant agreement ID n.o 101059532. This project was extended for 6 months by the Franziska Seidl Funding Program of the University of Vienna. LMV acknowledges support by the German Academic Scholarship Foundation (Studienstiftung des deutschen Volkes) and the Marianne-Plehn-Program of the Elite Network of Bavaria. KH acknowledges support by the Royal Society through a Dorothy Hodgkin Fellowship to KA Oman (DHF/R1/231105)
S.K.Y. acknowledges support from the Korean National Research Foundation (RS-2025-00514475; RS-2022-NR070872).

\newpage
\appendix
\section[\appendixname~\thesubsection]{Pressure gradient correction}
\label{sec:PGC}
%\subsection[\appendixname~\thesubsection]{}
Before disentangling the baryonic components and estimating the DM fraction from RCs, we applied the correction for pressure support on observed RCs, following the pressure gradient correction (PGC) method, introduced in  \cite[][Sec.~3.2]{GS21a}, defined as:
\begin{equation}
\label{eq:Vadc}
V^{PGC}_{c} = \sqrt{V^2_{\phi} - \sigma^2_R \Big[ \frac{\partial ln\Sigma}{\partial ln R} + \frac{\partial ln\sigma^2_R}{\partial ln R} + \frac{1}{2}(1-\alpha) \Big]} 
\end{equation}
where $V^{PGC}_{c}$ is the pressure gradient-corrected circular velocity, $V_{\phi}$ is the inclination-corrected rotation velocity, $\Sigma$ is the 2D mass surface density of the tracer (e.g., $\mathrm{H}\alpha$), and $\sigma_{R}$ is the radial velocity dispersion. Here, we follow \cite{Anne2008}, where the velocity anisotropy $(1- \sigma_\phi^2/\sigma^2_R)$ is approximated as $(1-\alpha)/2$, with $\alpha$ representing the logarithmic radial slope of the rotation velocity ($\alpha = \partial \ln V_\phi/ \partial \ln R$). The intrinsic kinematic profiles required for Equation~\ref{eq:Vadc} ($\Sigma$, $\sigma_R$, and $V_\phi$) are derived directly from the 3D forward kinematic modelling of the datacubes (discussed in Section~\ref{sec:Barolo}).

\begin{figure*}[h]
	\begin{center}
\includegraphics[angle=0,height=8.5truecm,width=12.5truecm]{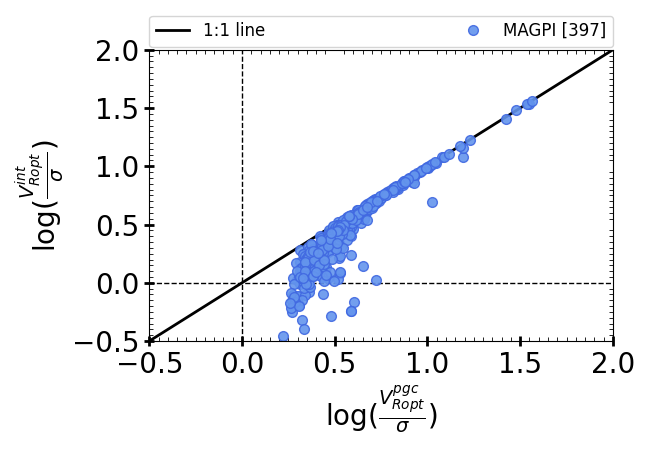} 					
    \caption{Comparison of the intrinsic and pressure-corrected rotation-to-dispersion ratios for the MAGPI galaxy sample. The $x$-axis shows the pressure-corrected ratio ($V_{\rm Rout}^{\rm PGC}/\sigma$), while the $y$-axis displays the intrinsic, uncorrected ratio ($V_{\rm Rout}^{\rm int}/\sigma$). Each point represents a single galaxy, and the solid black line indicates the 1:1 relation. This comparison demonstrates how galaxies that initially appear dispersion-dominated shift toward rotation-supported regimes after applying the pressure gradient correction.
}
\label{fig:PGS_plot}
\end{center}
\end{figure*}
It is important to note that all parameters in Equation~\ref{eq:Vadc} are radially dependent. Extracting these intrinsic profiles via \textsc{3DBarolo} mitigates beam smearing effects in $V_\phi(R)$ and, to a large extent, in $\sigma_R(R)$, thereby stabilizing the derived rotation-curve slope ($\alpha$) and logarithmic derivative of velocity dispersion ($\frac{\partial ln\sigma^2_R}{\partial ln R}$). However, $\Sigma(R)$ remains the most susceptible to resolution limits. Because \textsc{3DBarolo} does not perform a full deconvolution of the surface brightness distribution, the recovered flux remains spatially smoothed by the PSF. Consequently, the measured radial density gradients ($\partial \ln \Sigma / \partial \ln R$) are systematically flattened in the inner regions.

To calculate the pressure gradient, we adopt the empirical smoothing method of \cite{GS21a} rather than imposing a global analytical Sérsic model \citep[e.g.,][]{ManceraPina2026}. While forward-modeled, PSF-deconvolved analytical profiles provide mathematically ideal, noise-free gradients, we cannot utilize our existing ProFit models for this correction. Our ProFit models were derived from broad-band MUSE I-band images, meaning they represent the spatial distribution of the stellar continuum ($\Sigma_\star$). Stars typically exhibit a much steeper central concentration (a higher Sérsic index) than the ionized gas disk. Applying this steeper stellar density gradient to the gas kinematics physically misrepresents the local gas pressure support and would artificially over-inflate the corrected circular velocities in the inner regions.

Crucially, as detailed in \cite{GS21a}, the hydrodynamic pressure correction strictly requires the radial density gradient of the specific gaseous tracer ($\nabla \Sigma_{\rm gas}$). That is, it is essential to calculate this correction using the exact same emission line from which the rotation curve is derived (e.g., utilizing the observed H$\alpha$ surface brightness and dispersion profiles to correct the H$\alpha$ rotation curve). Different emission lines trace regions of varying local gas density and ionization parameters, meaning their spatial distributions are unique and not interchangeable with the bulk stellar continuum, as they can exhibit different kinematic behaviors—a point we demonstrate in Section~\ref{sec:emline-RCs} (see first example of velocity dispersion curve in Figure~\ref{fig:emline-RCs}). %\textbf{Therefore, we compute the derivative directly from the local radial bins of the observed gas tracer using our empirical polynomial approach.}

We acknowledge a fundamental limitation of this approach: because beam smearing inherently flattens the very inner ($R < \text{FWHM}_\text{PSF}$) observed density profile, our measured gradient is shallower than the true intrinsic gradient ($|\partial \ln \Sigma_\text{obs} / \partial \ln R| < |\partial \ln \Sigma_\text{int} / \partial \ln R|$). Consequently, our applied pressure correction represents a conservative lower limit, underestimating the true circular velocity in the innermost radial bins ($R \lesssim \text{FWHM}_\text{PSF}$). However, this local derivative rapidly converges with the intrinsic pressure gradient at larger radii, yielding robust kinematic corrections in the disk outskirts. To strictly insulate our final results from this central resolution limit, we evaluate the dark matter fractions exclusively at representative scale lengths that sit securely outside the beam-smeared region ($R > \text{FWHM}_\text{PSF}$).

In Figure~\ref{fig:PGS_plot}, we demonstrate the effect of the pressure gradient on the kinematics of galaxies in our MAGPI sample. We compare the intrinsic rotation-to-dispersion ratio ($V_{\rm Ropt}^{\rm int}/\sigma$), uncorrected for pressure support, to the pressure-corrected ratio ($V_{\rm Ropt}^{\rm PGC}/\sigma$). The data points show that many galaxies initially classified as dispersion-dominated (with low $V_{\rm Ropt}^{\rm int}/\sigma$) move closer to or above the 1:1 line after applying PGC, indicating that their true kinematics are rotation-supported. This confirms that the PGC method effectively accounts for the impact of turbulent pressure on rotation velocity, allowing inclusion of these galaxies in our analysis.

%%%%%%%%%%%%%%%%%%%
\section[\appendixname~\thesubsection]{Robustness of Global Kinematic Models to Asymmetric Emission}
\label{sec:appendix_asymmetry}

In a large kinematic survey of intermediate-redshift galaxies, global kinematic models must naturally account for some degree of asymmetric gas distribution driven by data quality constraints or intrinsic morphological irregularities. To validate our formal error budget—and specifically to ensure our uncertainties are not artificially underestimated due to spatial covariance between adjacent pixels (as discussed in Section~\ref{sec:errors})—we performed independent kinematic fits for the approaching and receding sides of our entire sample.

%\textbf{First, we compared the kinematic position angles and rotation velocities derived from these independent side-fits against our fiducial global models (which fit both sides simultaneously). We found excellent symmetric agreement between the independent arms and the global fits, confirming that our global kinematic axes and velocities are not systematically biased toward either side of the galaxy.}

\begin{figure}[h]
	\begin{center}
\includegraphics[angle=0,height=6.0truecm,width=15.5truecm]{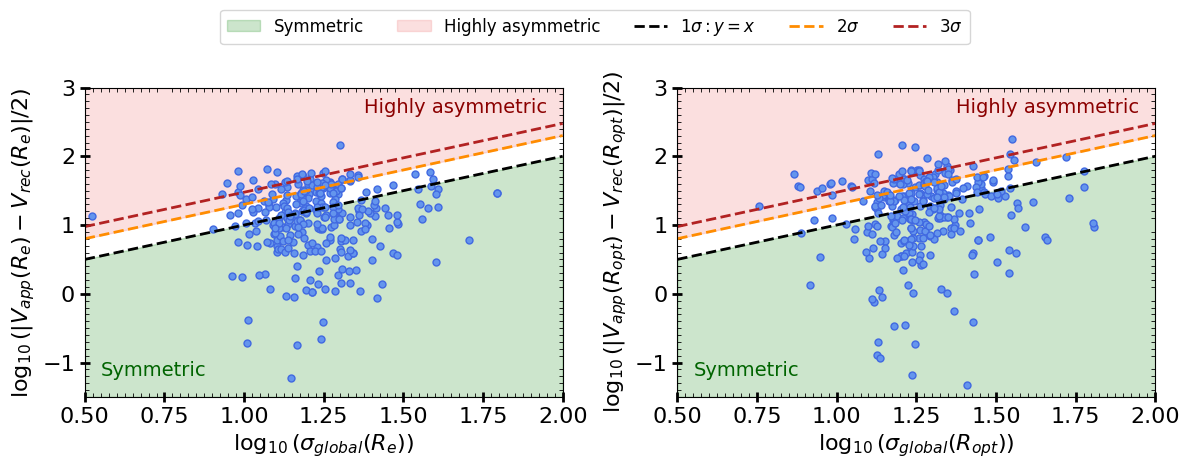} 
\caption{Diagnostic plot assessing whether the formal velocity uncertainties of the global kinematic models adequately capture the physical asymmetry of the galaxies. The kinematic asymmetry, defined as half the difference between the independently fitted approaching and receding velocities ($\Delta V = \vert{}V_{app} - V_{rec}\vert{} / 2$), is plotted against the formal global velocity uncertainty ($\sigma_{global}$) in logarithmic space. The left and right panels display these measurements evaluated at the effective radius ($R_e$) and the optical radius ($R_{opt}$), respectively. The dashed black, orange, and red lines denote the $1\sigma$ ($y=x$), $2\sigma$, and $3\sigma$ equivalence boundaries. The background is color-coded to distinguish kinematically symmetric systems (green shading; $\Delta V \le 2\sigma_{global}$) from highly asymmetric outliers (red shading; $\Delta V > 2\sigma_{global}$). The concentration of the sample below $2\sigma$ and within the green boundaries demonstrates that for the vast majority of galaxies, the global error budget robustly accounts for the differences between the approaching and receding sides.
}
\label{fig:BB_assymetry}
\end{center}
\end{figure}

We evaluated whether the formal global velocity uncertainties ($\sigma_{\rm glob}$) adequately capture the physical kinematic asymmetry, defined as the deviation between the independently fitted arms: $\Delta V = \vert V_{\rm app} - V_{\rm rec} \vert / 2$, where $V_{\rm app}$ and $V_{\rm rec}$ denote the rotation velocities derived from the independent fits of the approaching and receding arms, respectively.
We specifically evaluated this physical asymmetry using pressure-gradient-corrected velocities. The rationale is, local variations in gas density or turbulent velocity dispersion can induce apparent kinematic asymmetries in the raw observed rotation curve. By applying the pressure gradient correction first, we ensure that we are evaluating the true dynamical asymmetry of the galactic potential, independent of localized hydrodynamic effects.

In Figure~\ref{fig:BB_assymetry}, we plot this physical asymmetry against the formal global uncertainty at both $R_e$ and $R_{\rm opt}$. For the vast majority of the sample, the global error budget robustly accounts for the differences between the approaching and receding sides, with the data clustering securely below or within the $1\sigma$ and $2\sigma$ consistency boundaries (i.e., the formal uncertainty is larger than the physical asymmetry).

As expected, galaxies exhibit higher degrees of asymmetry in their outer, lower signal-to-noise regions. At $R_{\rm opt}$, 57 galaxies exhibit an asymmetry that exceeds the $2\sigma$ global uncertainty (compared to 50 galaxies at $R_e$). To ensure statistical transparency, these highly asymmetric outliers ($\Delta V > 2\sigma$) are explicitly flagged in our main kinematic catalog (released with this article). We conclude that our uniform global fitting approach is statistically robust, and the derived uncertainties reliably reflect the underlying structural asymmetries for the bulk of the sample.

%%%%%%%%%%%%%%%%%%%
\section[\appendixname~\thesubsection]{Surface brightness validation of stellar mass decomposition}
\label{sec:Mbulge-Mdisk}
We evaluate the robustness and structural validity of our constrained (fix-parameter) multi-component analytical stellar-mass decomposition methodology. As detailed in Section~\ref{sec:MModel}, our kinematic modeling relies on separating the global stellar mass profile into a rotationally supported, thin exponential disk ($n=1$) following a Freeman-disk \cite{Freeman}, and a central, pressure-supported spheroidal bulge modeled via a Hernquist density profile \cite{Hernquist1990}. Our methodology enforces physically motivated boundary conditions by fixing the total stellar mass budget to the integrated photometry ($M^{^{\rm SED}}_{\rm star, total} = M_{\rm disk} + M_{\rm bulge}$) and anchoring the sub-component geometries using established empirical scaling relations \citep[e.g.,][]{Morselli2017, Hon2023}. 
To verify that this physically constrained decomposition successfully reproduces the underlying stellar mass distribution without altering the global structural properties of the target galaxies, we cross-checked our multi-component models against the observed integrated single-Sérsic light distributions, where sersic index is estimated using photometry as discussed in Section~\ref{sec:data}. For each system, we reconstructed the independent 2D projected surface brightness profiles ($I = \frac{\Sigma}{\Upsilon}$). For the exponential disk and spherical bulge, this is analytically defined as:
%\begin{equation}%surface mass density (mass per unit area)
%\Sigma_{\rm disk}(R) = \frac{M_{\rm disk}}{2\pi R_{\rm disk}^2} \exp\left(-%\frac{R}{R_{\rm disk}}\right)
%\end{equation}
\begin{equation}%surface brightness (light per unit area)
    I_{\rm disk}(R) = \frac{M_{\rm disk}}{2\pi R_{\rm disk}^2 \Upsilon_{\rm disk}} \exp\left(-\frac{R}{R_{\rm disk}}\right)
\end{equation}
For the spheroidal bulge, the surface brightness is derived from the standard line-of-sight projection of the 3D Hernquist volume density, $\rho_{\rm bulge}(r)$:
\begin{equation}
    I_{\rm bulge}(R) = \frac{1}{\Upsilon_{\rm bulge}} \int_{-\infty}^{\infty} \rho_{\rm bulge}\left(\sqrt{R^2 + z^2}\right) dz
\end{equation}
where $\Upsilon$ represents the respective component mass-to-light ratios, which is assumed to be constant (M/L=1) in our case, i.e., $\Sigma = I$. We then evaluated whether the linear sum of these separate baryonic potentials closely traced the global photometric profile:
\begin{equation}
    \Sigma_{\rm total}(R) = \Sigma_{\rm disk}(R) + \Sigma_{\rm bulge}(R) \approx \Sigma_{\rm S\acute{e}rsic}(R)
\end{equation}

\begin{figure}[h]
	\begin{center}
\includegraphics[angle=0,height=6.0truecm,width=7.5truecm]{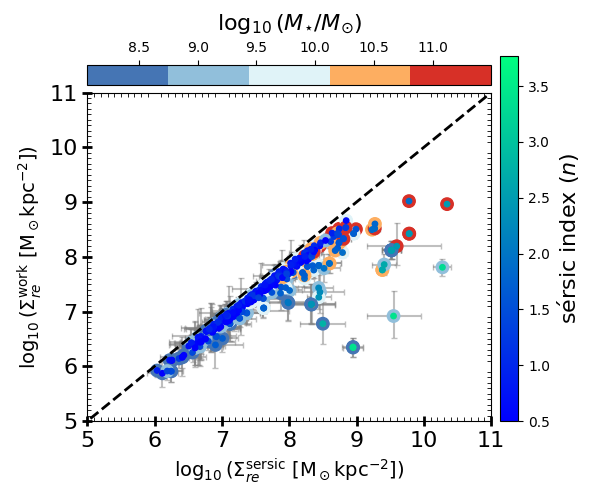} 	
\includegraphics[angle=0,height=6.0truecm,width=7.5truecm]{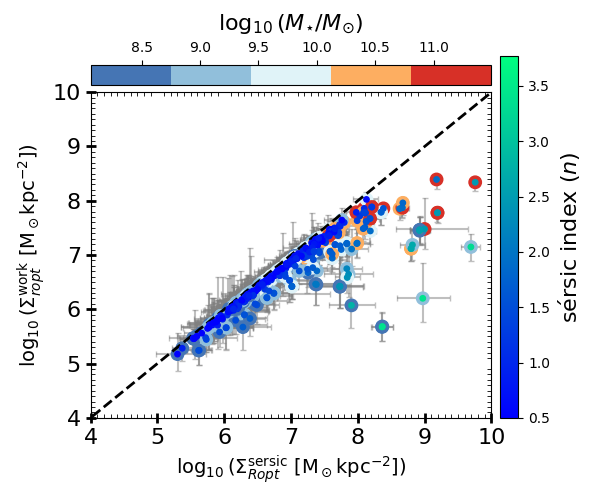} 	
\caption{surface-brightness obtained from an unconstrained, free-parameter MCMC optimization at $R_{\rm e}$ (left) and $R_{\rm opt}$ (right). In both panels, interior of data-points is color-coded by Sérsic index and exterior with total stellar mass.}
\label{fig:SB_comparison}
\end{center}
\end{figure}

We notice that representative multi-component surface mass density decompositions ($\Sigma_{\rm disk} + \Sigma_{\rm bulge}$) for both a low-mass and a high-mass system (a few exceptions), demonstrating excellent structural alignment with the empirical single-Sérsic light profile. The results are shown in Figure~\ref{fig:SB_comparison}, where we plot surface mass density evaluated at the effective radius ($R_{\rm e}$; left panel) and the optical radius ($R_{\rm opt}$; right panel). Majority ($\sim 95\%$) of sample shows a very tight one-to-one relation within $1\sigma$ uncertainty. However, we do notice large scatter in a few massive and high-sersic index galaxies, this is most-likely driven by localized structural variations (e.g., weak bars or star-forming clumps) or radial gradients in the stellar mass-to-light ratio. We stress that the `exponential disk + Hernquist bulge' framework is an underlying assumption of this study, chosen to systematically (uniformly across the sample) constrain the stellar mass distribution given the dataset's limitations, as detailed in Section~\ref{sec:errors}. 

We intentionally avoided unconstrained 2-component fitting (e.g., MCMC parameter optimization) because for seeing-limited intermediate-redshift data, free fitting suffers from severe parameter degeneracies (the ``bulge-disk conspiracy"). We noticed that unconstrained fits frequently unphysically trade mass between the bulge and inner disk to minimize global $\chi^2$, systematically underestimating the central surface brightness at all radii by up to $40-60\%$, which would artificially inflate the derived central dark matter fractions. Moreover, space-based imaging for the full MAGPI sample is not currently available at the depth required for uniform, unconstrained multi-component decomposition; however, a detailed multi-component structural analysis for a high-resolution sub-sample of $\sim30$ galaxies is currently being conducted by Mauro et al. (submitted).

Furthermore, we notice that some of the galaxies have Sérsic index below unity ($n < 1$). We argue that, while seeing-limited data systematically artificially suppresses the central density gradient—often driving the measured Sérsic index below unity ($n < 1$) in low-mass disks, and artificially lowering $n$ in massive, bulge-dominated systems—integral global parameters such as the effective radius ($R_{\rm e}$) and inclination remain highly robust as they are constrained by the extended outer disk. Consequently, anchoring our dynamical models to these global parameters, while explicitly fixing the disk component to a physical $n=1$ profile, effectively shields the inner mass distribution from central beam-smearing artefacts.

%%%%%%%%%%%%%%
\section[\appendixname~\thesubsection]{Gas masses}\label{sec:Mbar}
			Observations show that typical star-forming galaxies lie on a relatively tight, almost linear, redshift-dependent relation between their stellar mass and star formation rate, the so-called main sequence of star formation \citep[e.g., ][]{Noeske2007, Whitaker2012, speagle14}. Most stars since $z\sim 2.5$ were formed on and around this MS \citep[e.g., ][]{Rodighiero2011}, and galaxies that constitute it, usually exhibit a rotating disk morphology \citep[e.g., ][]{ForsterSchreiber2006, Daddi2010b, Wuyts2011b}. 
			Figure~\ref{fig:MSequence} shows the position of the final sample (detailed in Section~\ref{sec:Fsample}) with respect to the main sequence of typical star-forming galaxies (MS), i.e., their offset from the main sequence:
			\begin{equation}
				\delta {\rm MS} = {\rm SFR}/{\rm SFR}({\rm MS}; z,{M_{\rm star}}),
			\end{equation} 
			where ${\rm SFR}({\rm MS}; z,{M_{\rm star}})$ is the analytical prescription for the center of the MS as a function of redshift and stellar mass proposed in the compilation by \cite{speagle14}. This figure shows that majority of the galaxies in the total sample are within the $2-3\sigma$ range of SFMS. This enables us to estimate their molecular gas masses ($M_{\rm H2}$) using the \cite{Tacconi2018} scaling relations, which provide a parameterization of the molecular gas mass (including the Helium content) as a function of redshift, stellar mass, and offset from the MS stemming from a large sample of about 1400 sources on and around the MS in the range $z=0-4.5$ (see also \cite{Genzel2015} and \cite{Freundlich2019}). The scatter around these molecular gas scaling relations and the stellar mass induces a 0.3 dex uncertainty in the molecular gas mass estimates, which is accounted in the $f_{_{\rm DM}}$ error estimates.  The H2 mass of our sample is $7.75 \leq log(M_{\rm H2} \ [M_\odot]) \leq 10.18$, with an average molecular gas fractions ($f_{_{\rm H2}} = \frac{M_{\rm H2}}{M_{\rm bar}}$) of $0.10\pm 0.06$.

			To calculate the atomic mass ($M_{\rm HI}$) content of galaxies within the redshift range $0.1 \leq z \leq 0.85$, we use the HI scaling relation presented by \cite{Bianchetti2025} at $z=0.2-0.5$, and \cite{Chowdhury2022} at $z>0.5$. The relation was derived using a stacking analysis across three stellar mass bins, each bin with a $4\sigma$ detection and an average uncertainty of $\sim 0.3$ dex. This uncertainty in the gas scaling relation is additionally accounted in the error of HI mass estimates and propagated with MCMC when estimating the $f_{_{\rm DM}}$. The HI Mass range of our sample is $8.62 \leq log(M_{\rm HI} \ [M_\odot]) \leq 10.68$, with an average atomic gas fractions ($f_{_{\rm HI}} = \frac{M_{\rm HI}}{M_{\rm bar}}$) of $0.57\pm 0.19$. for further details of gas mass computation, we refer reader to \cite{GS23}.

%%%%%%%%%%%%%%%%%%%%%%%%%%%%%%%%%		
\section[\appendixname~\thesubsection]{Full-spectral fitting stellar masses}\label{sec:Mstar2}	
\begin{figure*}[h]
	\begin{center}
\includegraphics[angle=0,height=5.0truecm,width=7.5truecm]{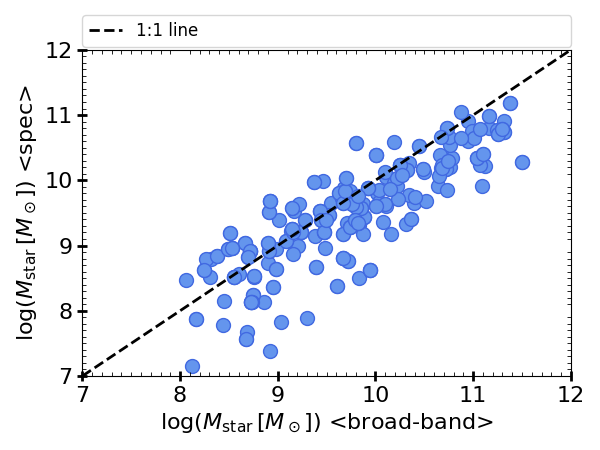} 	

\includegraphics[angle=0,height=5.0truecm,width=7.5truecm]{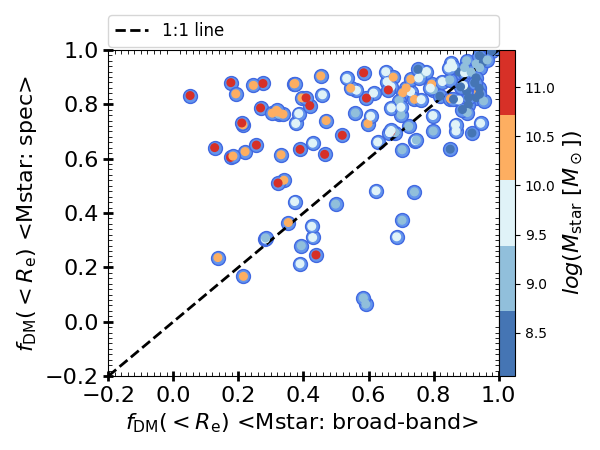}		
\includegraphics[angle=0,height=5.0truecm,width=7.5truecm]{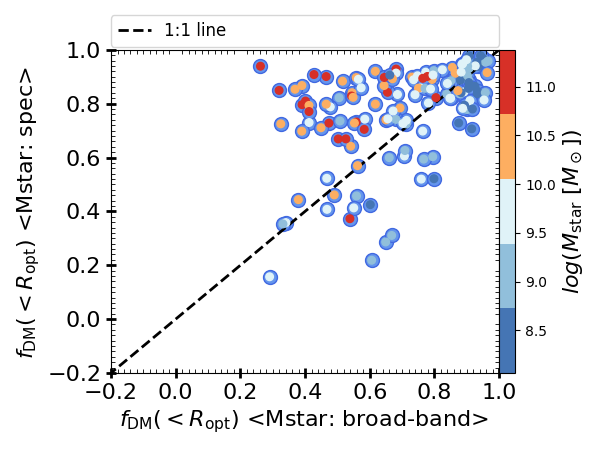}			
    \caption{Impact of stellar mass estimation techniques on DM Fraction measurements. {\em Top panel:}  Comparison between stellar masses derived from SED fitting using broad band images (used in this study; Bellstedt et al.) and those estimated using spatially resolved spectra-fitting techniques by Poci et al. (in preparation). {\em Bottom panel:} Comparison of DM fractions within $R_{\rm e}$ (left) and $R_{\rm opt}$ (right). The x-axis shows $f_{_{\rm DM}}$ computed using Bellstedt et al. stellar masses (broad-band), while the y-axis shows values based on Poci et al. estimates (spec). Data points are color-coded by the SED-derived stellar mass. The black dashed in all panels shows the one-to-one line for reference.}
\label{fig:Mstar2-fdm2}
\end{center}
\end{figure*}	

In the main analysis, stellar masses were estimated via broad-band photometric spectral energy distribution (SED) fitting, following the methodology outlined in \cite{Bellstedt2020}. While this technique yields reliable global mass estimates, we observe that a very small subset (n=3) of massive ($M_{star} > 10^{10}\ M_\odot$) galaxies in our sample exhibit unphysical (i.e., negative) DM fractions. Therefore, to assess the robustness of our DM fraction ($f_{_{\rm DM}}$) measurements, we compare them to estimates based on stellar masses derived using a full-spectral fitting technique (Poci et al., in preparation).

In this method, each MAGPI mini-cube is spatially-integrated to produce a single spectrum. For the subset of galaxies for which this spectrum has \(S/N>20\), we fit them in pPXF \citep{Cappellari2023} using the E-MILES SSP library \citep{Vazdekis2010} assuming a Chabrier IMF. From the resulting star-formation history, we compute the stellar mass-to-light ratio, and from the spectrum itself we compute the spatially-integrated luminosity over the same spectral window. This results in a spatially-integrated stellar mass.

Figure~\ref{fig:Mstar2-fdm2}, presents the resulting $f_{_{\rm DM}}$ values computed using these full-spectral stellar masses. We find that broad-band SED fitting based stellar masses are systematically higher than those derived via full-spectral fitting. Consequently, especially in high-mass systems, the DM fractions derived in this work maybe underestimated by more than 50\%—leading  to the appearance of ``dark matter-deficient ($f_{_{\rm DM}}<20\%$)" galaxies. In the low mass galaxies this difference is only within $\pm 5-10\%$. That is, applying full spectral-fitting based stellar masses significantly increases the inferred $f_{_{\rm DM}}$ within both $R_{\rm e}$ and $R_{\rm opt}$, reaffirming that even the most massive galaxies in our sample contain substantial DM  components. Nevertheless, as the full-spectral fitting stellar masses are not yet available for full sample, we adopt the photometric SED-based stellar masses as our baseline for all primary analyses presented in this study.

%\paragraph{Note added.} This is also a good position for notes added after the paper has been written.

% Bibliography

%% [A] Recommended: using JHEP.bst file
%\bibliographystyle{aa}
\bibliographystyle{JHEP}
\bibliography{DM_fraction_2025.bib}

%% or
%% [B] Manual formatting (see below)
%% (i) We suggest to always provide author, title and journal data or doi:
%% in short all the informations that clearly identify a document.
%% (ii) please avoid comments such as "For a review'', "For some examples",
%% "and references therein" or move them in the text. In general, please leave only references in the bibliography and move all
%% accessory text in footnotes.
%% (iii) Also, please have only one work for each \bibitem.

% \begin{thebibliography}{99}

% \bibitem{a}
% Author,
% \emph{Title},
% \emph{J. Abbrev.} {\bf vol} (year) pg.

% \bibitem{b}
% Author,
% \emph{Title},
% arxiv:1234.5678.

% \bibitem{c}
% Author,
% \emph{Title},
% Publisher (year).

% \end{thebibliography}

\end{document}